\documentclass[acmlarge]{acmart}
\usepackage[figuresright]{rotating}
\usepackage{graphicx}
\usepackage{enumerate}
\usepackage{caption}
\usepackage{algorithm}
\usepackage{algorithmic}
\usepackage{color}
\usepackage{subfigure}
\usepackage{graphicx}
\usepackage{stfloats}
\usepackage{float}
\usepackage{amsthm}
\usepackage{amsmath}
\usepackage{amsfonts}
\usepackage{booktabs}
\usepackage{multirow}

\usepackage{amssymb}
\usepackage{diagbox}
\usepackage{color}
\usepackage{pifont}
\usepackage{balance}
\usepackage{picins}
\usepackage[utf8]{inputenc}

\newtheorem{defi}{Definition}
\newtheorem{thm}{Theorem}
\AtBeginDocument{%
  }

\setcopyright{acmcopyright}
\copyrightyear{2022}
\acmYear{2022}
\acmDOI{XXXXXXX.XXXXXXX}

\acmJournal{JACM}
\acmVolume{37}
\acmNumber{4}
\acmArticle{111}
\acmMonth{8}

\begin{document}

\title{DPIVE: A Regionalized Location Obfuscation Scheme with Personalized Privacy Levels}

\author{Shun Zhang}
\affiliation{%
  \institution{Anhui University}
  \city{Hefei}
  \postcode{230601}
  \country{China}
}
\email{szhang@ahu.edu.cn}

\author{Pengfei Lan}
\affiliation{%
  \institution{Anhui University}
  \city{Hefei}
  \postcode{230601}
  \country{China}
}
\email{e21201061@stu.ahu.edu.cn}

\author{Benfei Duan}
\affiliation{%
  \institution{Anhui University}
  \city{Hefei}
  \postcode{230601}
  \country{China}
}
\email{dbf97@stu.ahu.edu.cn}

\author{Zhili Chen}
\affiliation{%
  \institution{East China Normal University}
  \city{Shanghai}
  \postcode{200062}
  \country{China}
}
\email{zhlchen@sei.ecnu.edu.cn}

\author{Hong Zhong}
\affiliation{%
  \institution{Anhui University}
  \city{Hefei}
  \postcode{230601}
  \country{China}
}
\email{zhongh@ahu.edu.cn}

\author{Neal N. Xiong}
\affiliation{%
  \institution{Sul Ross State University}
  \city{Alpine}
  \postcode{TX 79830}
  \country{USA}
}
\email{xiongnaixue@gmail.com, neal.xiong@sulross.edu}



\renewcommand{\shortauthors}{Zhang et al.}

\begin{abstract}

The popularity of cyber-physical systems is fueling the rapid growth of location-based services. This poses the risk of location privacy disclosure. Effective privacy preservation is foremost for various mobile applications. Recently, geo-indistinguishability and expected inference error are proposed for limiting location leakages. In this paper, we argue that personalization means regionalization for geo-indistinguishability, and we propose a regionalized location obfuscation mechanism called DPIVE with personalized utility sensitivities. This substantially corrects the differential and distortion privacy problem of PIVE framework proposed by Yu et al. on NDSS 2017. We develop DPIVE with two phases. In Phase I, we determine disjoint sets by partitioning all possible positions such that different locations in the same set share the Protection Location Set (PLS). In Phase II, we construct a probability distribution matrix in which the rows corresponding to the same PLS have their own sensitivity of utility (PLS diameter). Moreover, by designing QK-means algorithm for more search space in $2$-D space, we improve DPIVE with refined location partition and present fine-grained personalization, enabling each location to have its own privacy level endowed with a customized privacy budget. Experiments with two public datasets demonstrate that our mechanisms have the superior performance, typically on skewed locations.

\end{abstract}



\maketitle

\section{Introduction}\label{sec:introduction}

With the rapid development of smart sensing and cloud/fog computing, sensor networks has promoted the popularity of Cyber-Physical Systems (CPSs) that can achieve interconnection between the physical world and cyberspace. With CPS services, mobile users can sense their location and get some Location-Based Services (LBSs), such as Uber and Didi Chuxing.
In recent years, LBSs have achieved broadly public acceptance and adoption, and even play an indispensable role in people's livings.
With benefiting from LBSs, users' locations are continuously collected by untrusted service providers, which leads to the disclosure of location privacy, such as working place and habitation \cite{BOG19,LZF18}. Then the adversary can attack more sensitive information of the users based on their background knowledge. Therefore, how to protect user's location privacy in LBSs is an urgent problem to be solved \cite{WWL20,WXH12}.

Geo-indistinguishability \cite{ABC13} and expected inference error \cite{STL11,STT12} are two privacy notions recently used for location privacy protection. Geo-indistinguishability deriving from differential privacy ensures that for two arbitrary locations within a certain distance, their produced pseudo-locations are similarly distributed. Then, an adversary with any prior knowledge can not infer the true location by observing the pseudo-location. The expected inference error reflects the accuracy of the adversary to guess the true location by observing the pseudo-location and using available prior knowledge.

Since 2015 some authors \cite{Sho15,OTP17} have proposed that expected inference error and geo-indistinguishability can be combined to protect location privacy. Later, Yu et al. \cite{YLP17} formally study the relationship between the two privacy notions and verify that they are complementary. Indeed, geo-indistinguishability only limits the adversary's posterior knowledge after observing the pseudo-location, but does not consider the adversary's inference attack based on prior knowledge, such as the distance between the inferred and true location, while the expected inference error does not consider the constraint on the posterior information derived from the release of pseudo-locations. For this, they propose PIVE, a two-phase dynamic differential location privacy framework. In Phase I, it searches for the Protection Location Set (PLS) satisfying the privacy requirements on each (true) location, and in Phase II, it publishes the pseudo-location through the differential privacy mechanism. However, the PLS of each location depends on its local situation. Then, the PLSs generally have different diameters and even intersect with each other. Thus, the proof of differential privacy for PIVE is problematic with respect to geo-indistinguishability. Moreover, due to narrow guesses within the actual PLS, the condition introduced in PIVE is confirmed to be not sufficient for bounding expected inference errors from below. Our recent paper \cite{ZDC21} confirms these differential privacy problems and proposes a couple of correction approaches with analyzing theoretically their satisfied privacy characteristics. The constructive privacy framework is still left open.

To finish the problems in PIVE pointed above, we should ensure that all PLSs have the same diameter if any two of them have the possibility of intersecting with each other, or all PLSs can have different diameters if any two of them do not intersect with each other (which implies regionalization of PLSs).
Following the latter, we should address three challenges as follows: 1) satisfying the personalization of sensitivity and improving the data utility, 2) achieving the differential privacy inside each PLS, and 3) allowing for the scenario with skewed locations.

For this,we propose DPIVE a regionalized mechanism in this paper. Given the relevant privacy parameters, the set of entire locations involved is divided into multiple disjoint PLSs, and the locations in the same PLS share the same diameter. The global lower bound
of inference error is transferred to requirements on each PLS.
This approach ensures
the $\epsilon$-DP within each PLS and weak DP on the whole
domain.
 We first propose QK-means, a $2$-D method replacing the former approach based on $1$-D Hilbert curve for region partitioning, which is much helpful to reduce the service quality loss. Besides, we consider the more general scenario that allows users to personalize their privacy budget on each location, and we develop PDPIVE mechanism that meets the personalized requirements of location privacy.

This paper introduces regionalization in 2-D space to the task of location obfuscation. Our proposed regionalized framework DPIVE achieves differential privacy protection and its personalization PDPIVE satisfies user's specified privacy on each PLS level. The main contributions are as follows.
\begin{enumerate}[(1)]
\item We consider the scenario where the user wants to protect her/his true location by reporting a pseudo-location in a domain of discretized locations and may have potential requirements of geo-indistinguishability and expected inference error.
    For this, we propose DPIVE a privacy mechanism that utilizes regionalization of PLSs to personalize sensitivity while ensuring differential and distortion privacy level.

\item
We design the QK-means algorithm to expand the search space of partitions
for disjoint PLSs in the $2$-D space, which greatly improves the data utility.
As for the scenario with personalized privacy budget on each location, we develop PDPIVE a personalized obfuscation mechanism
    that divides the domain into more compact PLSs for smaller quality loss.

\item We carry out a series of experiments on two public datasets. The results demonstrate that, our DPIVE approach
saves up to $15.8\%$ quality loss compared to the existing mechanisms while achieving desired privacy protection on skewed locations, and PDPIVE exhibits higher quality of obfuscation.
\end{enumerate}

The remainder of this paper is structured as follows. In
Section 2, we conduct a survey of related work. Section 3
introduces some necessary backgrounds. Section 4 describes
the proposed privacy framework, provides the QK-mean clustering technique and designs
the personalized privacy framework. Experimental results
are presented in Section 5. Finally, we conclude this
paper in Section 6.

\section{Related Work}\label{sec:related work}
Due to rapid development of smart sensing and computing capacities, Cyber-Physical Systems (CPSs) have achieved unprecedented levels of performance and efficiency in many areas. In particular,
CPS employs Internet of Things (IoT) and Industrial IoT for automation of real-world duties, in which sensitive personal data are involved.
This expedites the issue of privacy threats as an important challenge in academic community
\cite{YXP13,KKS21}. Butun et al. \cite{BOG19} proposed a location privacy preserving scheme for the IoT users of CPSs. Liu et al. \cite{LZF18} presented an EPIC framework that includes a differentially private mechanism to defend smart homes against the traffic analysis attack. Hong et al. \cite{HWJ18} proposed an attacker location evaluation-based fake source scheduling method, which addressed the problem of scheduling fake sources to enhance source location privacy and maintained system performance.

The methods guaranteeing location privacy have been extensively studied in the past decade \cite{CEPP17}. Many techniques are proposed, such as cloak-region, dummy location, and cryptographic solutions.
Li et al. \cite{LZW19} proposed a novel privacy preserving LBS query scheme, which combined the $k$-anonymity technique, the pseudo random function, and the Paillier cryptosystem.

The notion \emph{$k$-anonymity} is the most widely used anonymous method for protecting location privacy in the literatures. This technique produces $k-1$  dummy locations to construct an anonymous domain, 
such that the attacker can not infer which is the real location among the set of $k$ locations \cite{HL04}.
However, one limitation of $k$-anonymity is that all users involved report their real location and are assumed to be trusted. Zhao et al. \cite{ZLZ18} proposed ILLIA which enables k-anonymity-based privacy preservation against location injection attacks in continuous LBS queries. In the meantime, Jiang et al. \cite{JZW18} presented RobLoP, a robust location privacy preserving algorithm against location-dependent attacks.
Wang et al. \cite{WXH12} formalized an optimization problem for cloaking area generation, which utilizes users' footprints to decide the
cloaking areas with privacy requirements expressed through both $k$-anonymity and entropy based
metrics.
However, only using anonymous method can not achieve good protection to a wide range of data and is vulnerable to background knowledge attack \cite{YXS18}. Homomorphic encryption \cite{ABL20} is a good model to ensure the confidentiality of task’s location policy \cite{YLL19} but induces extra computational cost, and the availability of data decreases greatly \cite{YXS18,ABC13}.

\emph{Expected inference error} is a stronger privacy notion first proposed by Shokri et al. \cite{STL11}, which is a natural way to measure the location privacy  by the expected distance between the guessed location
by the adversary and the real location. Then a number of location
obfuscation mechanisms have been developed relying on this notion.
 In \cite{STT12}, an optimal obfuscation mechanism for achieving maximum level of privacy was designed by solving a linear program with constraint on the service quality loss.
Ahmad et al. \cite{Ahmad18} developed an effective intent-aware query obfuscation solution to maintain Bayes-Optimal Privacy in a personalized web search environment.
 The expected inference error can resist against the Bayesian attack to some extent, however, it does not take into account the constraint on the posterior information gain obtained by the reported pseudo-locations \cite{YLP17}.

\emph{Differential Privacy (DP)} \cite{Dw06} has emerged as the \emph{de facto} standard privacy notion for privacy-preservation research on data analysis and publishing.
Andres et al. \cite{ABC13} introduced \emph{geo-indistinguishability}, a strong concept based on differential privacy, which ensures that any two geographically close locations have similar probability distributions on any pseudo-location so that the adversary can not infer the true location by observing the pseudo-locations. Due to this, several location privacy protection mechanisms have been proposed recently \cite{XWW20,TTZ20,BCP14,Sho15,WWL20,RT20}. Wu et al. \cite{WWL20} proposed a location privacy-preserving system for LBS, which constructed high-quality ``cover-up ranges" to make it difficult for an attacker on the untrusted server-side to learn users' query locations or query ranges. Xu et al. \cite{XWW20} proposed a geo-indistinguishability based framework to preserve the privacy of individuals on ride-sharing platforms. Ren et al. \cite{RT20} presented a vehicle location privacy protection framework called Expanding Geo-Indistinguishability framework (EGeoIndis). Tao et al. \cite{TTZ20} investigated privacy protection for online task assignment with the objective of minimizing the total travel distance.

The scheme in \cite{BCP14} used linear programming to minimize global expected service quality loss averaged over all locations, with a uniform privacy parameter for geo-indistinguishability.
Later, Some authors \cite{Sho15, OTP17} proposed to combine the two privacy notions using linear programming. Qiu et al. \cite{QSP22} designed a location obfuscation strategy to minimize the quality-of-service loss of task distribution without compromising workers’ location privacy. For further scenario applications, several mechanisms are applied in mobile crowdsourcing for optimal task allocation \cite{WYH19,HLZ20,WWY22}. Zhang et al. \cite{ZZX21} proposed two novel privacy-preserving task recommendation schemes for mobile crowd sensing.
Niu et al. \cite{NCW20} proposed Eclipse, which is a three-phase differential location privacy-preserving mechanism by using PIVE \cite{YLP17}, to effectively prevent mobile user’s location privacy from the long-term observation attacks. Gursoy et al. \cite{GLT18} presented DP-Star, a methodical framework for publishing trajectory data with differential privacy guarantee as well as high utility preservation.


Recently,
Yu et al. \cite{YLP17} pointed out that the formulation above \cite{Sho15,STT12,BCP14,WYH19}
uses uniform differential privacy parameter and emphasizes the globally average performance on privacy/quality metrics over all locations. For this, they formally examined the relationship between the two privacy notions and propose PIVE mechanism with adding user-defined lower bound of inference error. PIVE is a two-phase dynamic differential location privacy framework that focuses on local performance of privacy protection. In phase I, it searches for the Protection Location Set (PLS) satisfying user's privacy requirements for the true location, and in phase II, it publishes the pseudo-location through  the exponential mechanism. However, we found that PIVE fails to provide provable privacy guarantee on adaptive protection location sets as claimed, and we discussed this problematic framework in detail in \cite{ZDC21}. In short, the diameter of the PLS obtained in PIVE by adaptive search around each apriori location is generally different and there exist intersection cases for PLSs, which leads to that PIVE can not theoretically preserve differential privacy on the PLSs. We also proposed a pair of possible correction approaches and analyze their respective privacy characteristics. Particularly, the results on geo-indistinguishability (or differential privacy) within each region and over more general regions are presented therein.

In this paper, we are intended to correct the problematic construction of PIVE. Given the relevant privacy parameters and conditions, the entire location set is partitioned into multiple disjoint parts. Each part is assigned as the PLS for all apriori locations inside and ensures the lower bound of the inference error. Thus, the locations within the same PLS are protected with strong differential privacy, while those across different PLSs protected with weak differential privacy. Our proposed DPIVE mechanism allows users to define their own privacy level for both phases. Besides, for the personalization of privacy budget at each location, we implement the location obfuscation mechanism PDPIVE theoretically and practically.

\section{System model and definitions}

In this section we first introduce the notation of geo-indistinguishability and differential location privacy, describe the model of the adversary model used in this paper. Then, we present the problem to be addressed in this paper. Table \ref{tlb:notation} summarizes the notations used in our work.

\begin{table} \small
\caption{Summary of Notations}
\label{tlb:notation}
        \centering
		\begin{tabular}{ll}
			\toprule
			Symbol  & \quad\quad\quad\quad\quad\quad\quad Definition \\
			\midrule
                        $\epsilon_0,\ \epsilon_k$      & Total privacy budget and privacy level on $\Phi_{k}$ \\
                        $\mathcal{X}$ &  Set of the user's possible locations
\\
						$f(x'|x)$       & Probability of reporting location $x'$ for the actual $x$
\\
                        $d(x,y)$        & Travel distance between the locations $x$ and $y$
                        \\
                        $\epsilon_g,\ \theta$ & Geo-indistinguishability parameter and its deviation
                        \\
                        $\Phi$    & Protection Location Set (PLS)
                        \\
                        $D(\Phi)$ & Diameter of $\Phi$ (the largest distance between two points inside)
                        \\
                        $\Delta q$   & Sensitivity of the scoring function $q$
                        \\
                        $\pi$         & Prior probability
                        \\
                        $ExpEr(x')$    & Conditional expected inference error for reported $x'$
                        \\
                        $E_m$         & Minimum (local) inference error
                        \\
                        $\hat{x}$    & The location estimated by optimal inference attack
                        \\
                        ExpErr        & Unconditional expected inference error
                        \\
                        QLoss          & Service quality loss
                        \\
                        $\Delta u(\Phi_k)$     & The sensitivity of $u$ on PLS $\Phi_k$
                        \\
                        $\mathcal{K}$    & Exponential Mechanism
                        \\
                        $AvgErr(x)$   &  Average inference error of optimal inference attack for $x$
                        \\
                        $p_s$   &  Success probability of Bayesian inference attack
                        \\
			\bottomrule
		\end{tabular}
	\end{table}

\subsection{Differential Location Privacy}

Differential Privacy (DP) \cite{Dw06} is a strict privacy concept that provides provable privacy protection for users. Regardless of the adversary's prior knowledge, it ensures that any adversary can not determine the presence of a particular individual from the processed data set.
Geo-indistinguishability based on differential privacy \cite{ABC13} is a statistical notion of location privacy, which has been widely used in the field of location privacy protection. To achieve DP protection
over PLS, we use the loose definition as follows.

\begin{defi}[$(\epsilon_g,\theta)$-Geo-indistinguishability within PLS \cite{ZDC21}]\label{def:geoI}
Assume that the probability distribution $f(\cdot|\cdot)$ for a mechanism $\mathcal{A}$ satisfies, for any $x, y$ in PLS $\Phi\subset\mathcal{X}$,
\begin{equation}
\frac{f(x^\prime|x)}{f(x^\prime|y)}\leq  e^{\epsilon_g \left(d(x,y)+\theta\right)},\ \ \ \  x^\prime\in \mathcal{X},
\end{equation}
then $\mathcal{A}$ is $(\epsilon_g,\theta)$-geo-indistinguishable on $\Phi$. If $\theta=0$, we say that $\mathcal{A}$ gives $\epsilon_g$-geo-indistinguishability on $\Phi$ without deviation.
\end{defi}

This means that two geographically close locations have similar probability distributions, which theoretically achieves that they are indistinguishable to each other for the adversary. Here, $\epsilon_g$ represents the geo-indistinguishability parameter that is determined by the privacy budget and the circular region usually centered at the user's location. All locations in the region have similar release distribution $f$ so that the true location can be hidden in this region, and the whole locations in this region are called the Protection Location Set (PLS). Accordingly, differentially
private location obfuscation can be defined as follows.

\begin{defi}[Local DP on PLS \cite{YLP17,ZDC21}]\label{def:eps_DPPLS}
A randomized location obfuscation mechanism
$f(\cdot|\cdot)$ achieves $\epsilon$-differential privacy on protection location
set $\Phi$, if for any locations $x, y \in\Phi$, and any output $x'\in\mathcal{X}$, we have
\begin{equation}\label{defi:eps-DP}
\frac{f(x^\prime|x)}{f(x^\prime|y)}
\leq  e^{\epsilon}.
\end{equation}
\end{defi}

For functions where the output space is non-numeric, the exponential mechanism
is widely used to achieve differential privacy. It requires a scoring function $q:\ \Phi\times\mathcal{X}\rightarrow \mathbb{R}$ which assigns a real-valued score to each point-point pair, ideally such that each $x'\in\mathcal{X}$ with good utility receives a high score. Due to the PLS scenario,
two locations are regarded to be neighboring to each other if in the same PLS.

\begin{defi}[Sensitivity on PLS \cite{DR14}]\label{def:Sensitivity}
Let $x_1, x_2$ be any pair of neighboring locations (in PLS $\Phi$) and $x'\in\mathcal{X}$. The sensitivity of the scoring function $q$ on $\Phi$ is given by, its maximal change,
\begin{equation}
\Delta q = \sup_{x_1,\, x_2,\, x'} \left| {q(x_1,x') - q(x_2,x')} \right|.
\end{equation}
\end{defi}

\begin{defi}[Exponential Mechanism on PLS \cite{MT07,DR14}]\label{def:Exponential}
Given a scoring function $q$ on $\Phi\times\mathcal{X}$,
 the exponential mechanism
$\mathcal{M}(x,q)$ outputs $x' \in \mathcal{X}$  with probability proportional to $\exp \left(\frac{\epsilon q(x,x')}{2\Delta q}\right)$.
\end{defi}




\subsection{Bayesian Adversary Model}

 As all the Location-Based Service (LBS) providers require the access
permission to users' location data, the location privacy is  potentially disclosed to untrusted entities.
Knowing user's locations, an adversary can perform a broad spectrum of attacks. Thus, ensuring location privacy is foremost for LBS applications.


In LBS, users usually send their true locations to the service provider to get services. However, the service provider is often an untrusted entity and may disclose users' location privacy. For this, a common method is location perturbation, which generates a pseudo-location based on the true location and the user sends it to the server.

Following \cite{BCP14,HBM17,YLP17}, we suppose that the discretized location set $\mathcal{X}$ represents the user's possible locations.
An obfuscation mechanism takes the user's real location $x$ from $A$ as input and randomly chooses a
pseudo-location $x^\prime$ from $O$ with the probability distribution $f(x^\prime|x)$:
\begin{equation}
f(x^\prime|x)=\text{Pr}(O=x^\prime|A=x), \ \ \ \ \ x,\ x^\prime\in \mathcal{X}.
\end{equation}
In general, the objective of obfuscation mechanisms is mainly to design suitable probability distribution $f(\cdot|\cdot)$ in the sense of some metrics.

 As before \cite{MK12,STT12,YLP17}, we assume that the adversary has prior knowledge about user's location, which can be regarded as background knowledge to perform inference attacks.
 The adversary usually collects background knowledge by building a prior probability distribution  $\pi$  on  $\mathcal{X}$.
 The prior probability $\pi$ can be obtained via population density, historical locations and so on.
The adversary is also informed of the location obfuscation mechanism $f$. Assuming more information known by the adversary implies the higher privacy security of the required framework.

In the current scenario, the adversary infers the user's real location $x$ under the Bayesian adversary model.
  After the user reports her/his pseudo-location $x'\in\mathcal{X}$, the adversary computes the probability that each apriori location $x\in\mathcal{X}$ is the true location in the condition of generating $x'$, i.e., the posterior probability distribution $\text{Pr}(x|x')$, by

\begin{equation}\label{formu:post-dist}
\text{Pr}(x|x')=\frac{\text{Pr}(x,x')}{\text{Pr}(x')} =\frac{\pi(x)f(x^\prime|x)}{\sum_{x\in \mathcal{X}}\pi(x)f(x^\prime|x)}.
\end{equation}

Afterwards, a Bayesian adversary can launch an \textbf{optimal inference attack} to get the estimated location $\hat{x}$ which has the minimal expected inference error, i.e.,

\begin{equation}\label{eq:attack x1}
\hat{x}=\mathop{\arg\min}\limits_{y\in \mathcal{X}}\sum_{x\in \mathcal{X}}\text{Pr}(x|x^\prime)d_p(y,x),
\end{equation}
where $d_p$ is usually Euclidean distance $d$. When $d_p$ denotes Hamming distance $d_h$, that is, $d_h(x,x^\prime)=0$ if $x=x^\prime$, and $d_h(x,x^\prime)=1$ otherwise, this attack is called \textbf{Bayesian inference attack} and simply


\begin{equation}\label{eq:attack x2}
\hat{x}=\mathop{\arg\max}\limits_{x\in \mathcal{X}}\text{Pr}(x|x').
\end{equation}


 In such a scenario with Bayesian adversary attacks, the location privacy of a scheme can be measured by unconditional expected inference error \cite{STL11,STT12}, which is the expected inference error of adversary averaged on $\mathcal{X}$,

\begin{equation}\nonumber
ExpErr=\sum_{x'\in \mathcal{X}} \text{Pr}(x') \mathop{\min}\limits_{\hat{x}\in \mathcal{X}}\sum_{x\in \mathcal{X}} \text{Pr}(x|x')d(\hat{x},x)\quad \ \quad
\end{equation}

\begin{equation}
=\sum_{x'\in \mathcal{X}}\mathop{\min}\limits_{\hat{x}\in \mathcal{X}}\sum_{x\in \mathcal{X}}\pi (x)f(x'|x)d(\hat{x},x).
\end{equation}

The service quality loss is usually defined by the unconditional expected distance between true and perturbed locations,
\begin{equation}
QLoss=\sum_{x\in \mathcal{X}}\sum_{x'\in \mathcal{X}}\pi (x)f(x'|x)d(x',x),
\end{equation}
where the quality metric $d$ denotes the Euclidean distance as \cite{BCP14,Sho15}.

\subsection{Problem Statement}

\begin{figure}[tb]
\begin{minipage}[t]{1.0\linewidth}
\centering
\includegraphics[scale=0.3]{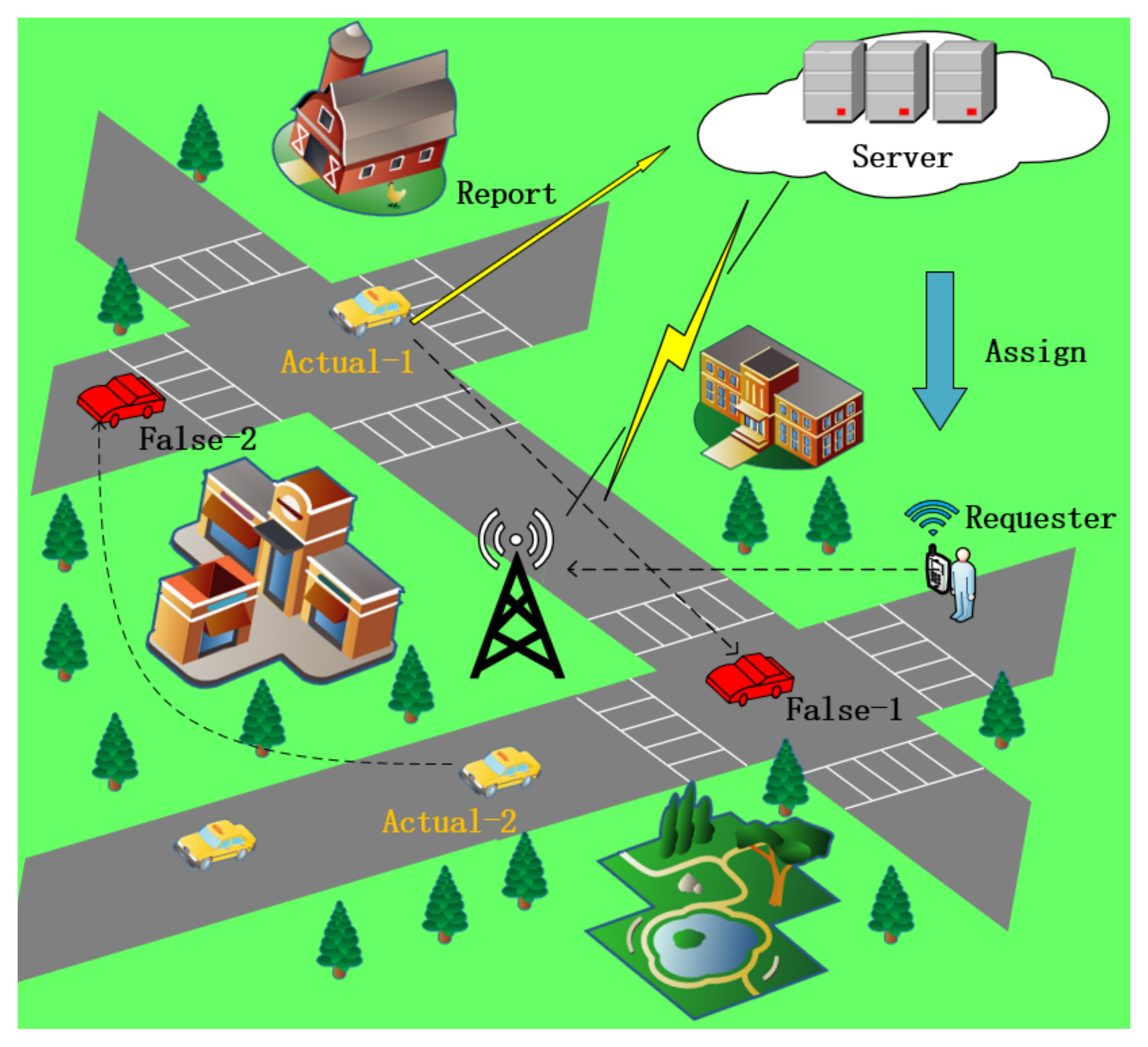}
\caption{A common scenario of location privacy in LBSs.}
\label{fig:scene}
\end{minipage}%
\end{figure}

In the mobile Internet era, users often have to report their real-time location for Location-Based Services (LBSs) while preserving their location privacy. Fig. \ref{fig:scene} shows a common location privacy scenario, which is common in the context of CPS. The users are located actually in ``Actual-1", ``Actual-2", etc., while reporting false positions accordingly labelled by ``False-1" and ``False-2", etc. Afterwards, the platform will assign services or tasks according to the reported false locations. In this scenario, a semi-trusted server
gathers data from mobile individuals and will faithfully process as required according to the gathered data.
 The collected data may be then aggregated
and continuously shared with some other untrusted entities for various purposes. This poses the issue how to generate a perturbed location at each user's side for reporting with location privacy guarantees.


Expected inference error and geo-indistinguishability are two statistical quantification based privacy notions. They can be integrated for globally optimizing utility subject to their joint guarantee \cite{Sho15,OTP17}. Later, they are argued to be complementary for location privacy and are combined effectively by developing PIVE, a two-phase dynamic differential location privacy framework \cite{YLP17}. Pseudo-locations (i.e., perturbed locations) are generated by exponential mechanism for achieving differential privacy over the PLS. However, the privacy framework turns out to be theoretically problematic, as pointed in our recent work \cite{ZDC21}. That is, in the given scenario the PLSs adaptively determined usually intersect with each other and each apriori location may have different diameters of PLSs, which directly harms the differential privacy preservation of the whole PIVE. To be worse, the condition for lower bound of inference errors is wrong because of the assumption of narrow
guesses within the actual PLS.
For this, we are intended to correct the location privacy model.

That is, under the same assumption as before that the user wants to protect the privacy of her/his true location by reporting a pseudo-location in a set $\mathcal{X}$ of nearby discrete locations.
 It is desirable to develop a location obfuscation mechanism that combines the two privacy notions and generates perturbed locations with effective local performance. The mechanism should allow that the informed adversary has prior knowledge of probability distribution $\pi$ over a discretized set $\mathcal{X}$ with the true location included and knows the location obfuscation distribution $f$.
Specifically, given the user's location,
 construct PLSs to make different apriori locations inside the same PLS share the same sensitivity (diameter) in the public mechanism, with preserving differential privacy. This motivates the presentation of DPIVE, a regionalized location privacy framework integrating both notions of location privacy.

Besides, realizing the personalization on user-controlled privacy budget enables mobile users to endow freely all locations with different privacy levels. How to optimize obfuscation mechanism from various perspectives (particularly to achieve smaller service quality loss) with respect to region partitioning is also a meaningful problem. To solve this, we develop PDPIVE a personalized framework together with quasi $k$-means clustering algorithm.

\section{Our Proposed DPIVE Scheme}\label{sec:our approach}

In this section we introduce DPIVE, a two-phase dynamic regionalization mechanism to protect location privacy including both geo-indistinguishability and expected inference error. We first propose the framework and then describe its two phases, partitioning Protection Location Sets (PLSs) and applying exponential mechanism with regionalized sensitivity, in detail. In the first phase, the core of our scheme, the set of discretized locations is partitioned into disjoint subsets (i.e., private PLSs) to protect user's true location, with preserving the expected location inference errors exceeding the user-defined lower bound against adversary's attacks via prior knowledge on the user's location. We develop a partitioning method of location set over a Hilbert curve selected optimally for determining disjoint PLSs. In the second phase, we utilize an exponential mechanism to generate pseudo-locations with small service quality loss, which produces a distribution matrix satisfying 1) independence of the input of true location, and 2) user's location privacy preferences on $\epsilon$ and $E_m$. Then, we prove the differential privacy for locations both within each PLS and across all PLSs.

\subsection{DPIVE Regionalization Framework}\label{}

Yu et al. \cite{YLP17} verify that geo-indistinguishability and expected inference error are two complementary notions, and recently Zhang et al. \cite{ZDC21} confirm a sufficient condition \eqref{eq:inequality} to ensure the lower bound on expected inference error.

As before ,  the conditional expected
inference error is
\begin{equation}
ExpEr(x')=\mathop{\min}\limits_{\hat{x}\in \mathcal{X}}\sum_{x\in \mathcal{X}}\text{Pr}(x|x')d(\hat{x},x), \ \ \text{for}\ x'\in \mathcal{X}.
\end{equation}



Let  $z=\mathop{\rm argmin}\limits_{\hat{x}\in \mathcal{X}}\sum_{x\in \mathcal{X}}\text{Pr}(x|x')d(\hat{x},x)$
 and denote $\text{Pr}(\Phi_k|x') = \sum_{y\in \Phi_k}\text{Pr}(y|x')$.
By normalization in each PLS $\Phi_k$ (with $\epsilon$-DP) from a partition $\{\Phi_k\}$,  we have
\begin{equation}\label{lower-bound-sum}
\begin{split}
ExpEr(x')=&\sum_{x\in \mathcal{X}}\text{Pr}(x|x')d(z,x)
\ge \sum_{k}\mathop{\min}\limits_{\widehat{x}_k\in \mathcal{X}}\sum_{x\in \Phi_k}\text{Pr}(x|x')d(\widehat{x}_k,x)
= \sum_{k} \text{Pr}(\Phi_k|x') \mathop{\min}\limits_{\widehat{x}_k\in \mathcal{X}}\sum_{x\in \Phi_k}\frac{\text{Pr}(x|x')d(\widehat{x}_k,x)}{\sum_{y\in \Phi_k}\text{Pr}(y|x')}\\
=& 
\sum_{k} \text{Pr}(\Phi_k|x') \mathop{\min}\limits_{\widehat{x}_k\in \mathcal{X}}\sum_{x\in \Phi_k}\frac{\pi(x)f(x'|x)d(\widehat{x}_k,x)}{\sum_{y\in \Phi_k}\pi(y)f(x'|y)}
\ge  
\sum_{k} \text{Pr}(\Phi_k|x') e^{-\epsilon}  E'(\Phi_k),
\end{split}
\end{equation}
\noindent

\noindent 
where
\begin{equation}
E'(\Phi)=\mathop{\min}\limits_{\hat{x}\in \mathcal{X}}\sum_{x\in \Phi}\frac{\pi(x)}{\sum_{y\in \Phi}\pi(y)}d(\hat{x},x).
\end{equation}
Since $\sum_{k} \text{Pr}(\Phi_k|x')=1$, 
the condition
that for all $\Phi_{k}$,
\begin{equation}\label{ndss31-pri}
E'(\Phi_k)\ge e^{\epsilon} E_m,
\end{equation}
implies the user-defined error threshold, $ExpEr(x^\prime)\ge E_m$, for the optimal inference attack using any observed pseudo-location $x'$.

\begin{thm}[\cite{ZDC21}]\label{thm:DPIVE}
Given a domain partition $\{\Phi_k\}$ and an observed pseudo-location $x^\prime$ in $\mathcal{X}$, suppose an obfuscation mechanism satisfies $\epsilon$-DP on each PLS $\Phi_k$. If $E'(\Phi_k)\geq e^{\epsilon}E_m$ for each $\Phi_k$, then
 $ExpEr(x^\prime)\ge E_m$ for the optimal inference attack.
\end{thm}

We mention that a similar assertion is given in \cite{YLP17} (Theorem 1). That is, the sufficient condition \eqref{ndss31-pri} is replaced by \begin{equation}\label{eq:inequality}
E(\Phi)\ge e^{\epsilon} E_m,
\end{equation}
 in \cite{YLP17}, where
\begin{equation}\label{ndss-EPhi}
E(\Phi)=\mathop{\min}\limits_{\hat{x}\in \Phi}\sum_{x\in \Phi}\frac{\pi(x)}{\sum_{y\in \Phi}\pi(y)}d(\hat{x},x).
\end{equation}
It is claimed in \cite{YLP17} that, given $\Phi$ is convex in the discrete set $\mathcal{X}$, the authors obtain $E(\Phi)=E'(\Phi)$. However, this is not true in general, and we present a counterexample as follows.

\begin{figure}[tb]
\begin{minipage}[t]{0.4\linewidth}
\centering
\includegraphics[scale=0.6]{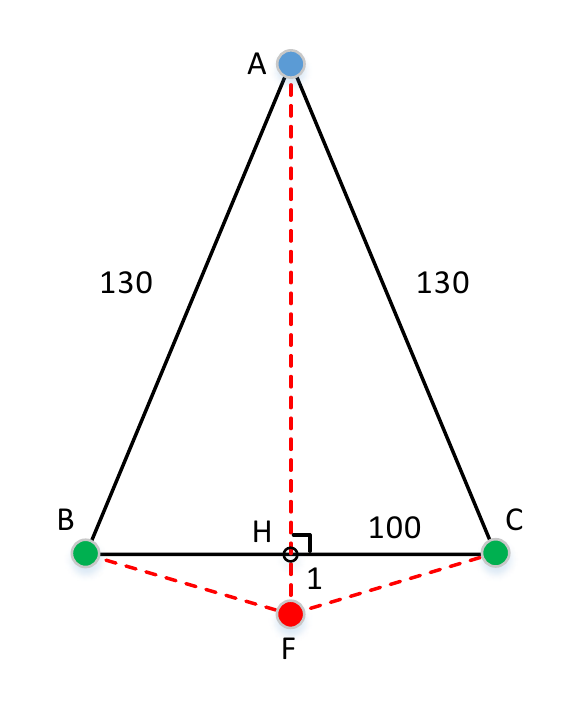}
\caption{Counterexample for convex PLS.}
\label{fig:convex}
\end{minipage}%
\begin{minipage}[t]{0.6\linewidth}
\centering
\includegraphics[scale=0.35]{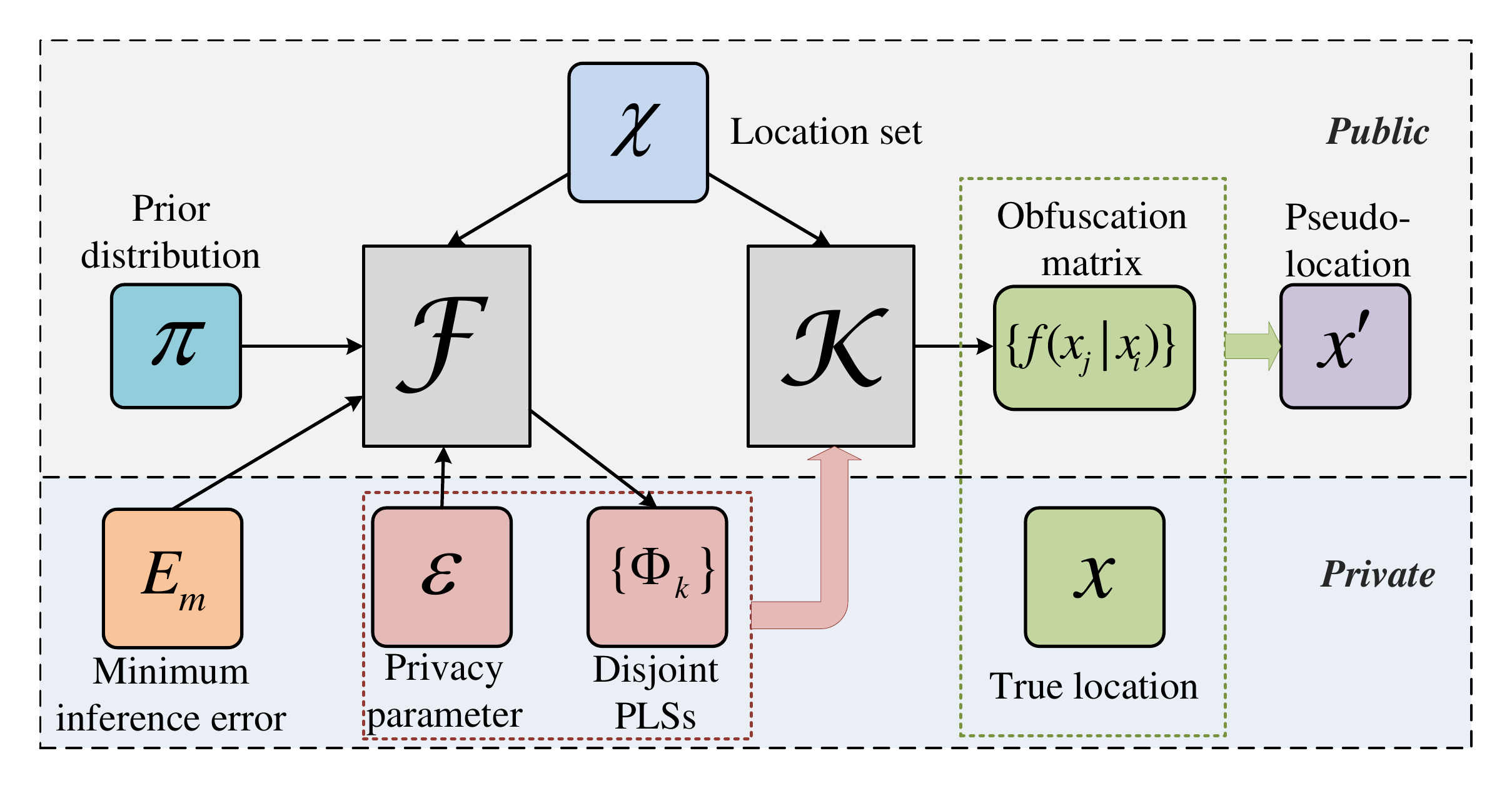}
\caption{The framework of DPIVE.}
\label{fig:DPIVE_model}
\end{minipage}%
\end{figure}

Suppose that, the prior distribution $\pi$ is uniformly distributed on $\mathcal{X}=\{A,B,C,F\}$, and $\Phi=\{A,B,C\}$, see Fig. \ref{fig:convex}. Obviously, $\Phi$ is convex in $\mathcal{X}$, that is, on the plane the convex hull of $\Phi$, the triangular range $\Delta ABC$ (the lengths of edges are $130, 130, 100$), does not include any point from $\mathcal{X}\backslash\Phi$. Then $E(\Phi)=76.7$ is larger than $E'(\Phi)=74.3$ since the minimal point for $E'(\Phi)$ is $F$ out of the range $\Delta ABC$.


The adaptive PLS for each apriori location is constructed based on the computation of \eqref{ndss-EPhi} in the phase I of PIVE.
Since $E(\{x_i\})=0$ for any single-point set, each PLS includes at least two locations. For each apriori location $x$, PIVE first searches in a large range for all possible sets of locations neighboring on Hilbert curve ranking that satisfy \eqref{eq:inequality} and PIVE chooses the set having the smallest diameter as PLS. Then in phase II, the diameter is assigned as the sensitivity of the exponential mechanism to generate pseudo-locations.

Unfortunately, the PLS obtained by PIVE depends locally on the true location adaptively and is usually different for each apriori location. Different PLSs may intersect with each other.
Then in the location obfuscation distribution matrix $\{f(x_j|x_i)\}$, each apriori location $x_i$'s row may have different sensitivities depending on the true location.  Such a problematic approach affects the differential privacy preservation on each PLS. We will review the PIVE Framework in Section \ref{subsect:PIVE}, see our paper \cite{ZDC21} for detailed analysis.

To solve this, we propose DPIVE, a regionalized location obfuscation mechanism. Given the privacy parameters without the input of true position, we first partition the entire discrete location set into $k$ parts, as many as possible each of which satisfies \eqref{ndss31-pri}. Then in the second phase each apriori location (row $i$) in the same part shares an identical sensitivity in exponential mechanism while all parts are regarded as possible PLSs symmetrically in the public location obfuscation distribution matrix. This means that any two apriori locations from different parts have no intersection on their PLSs and their corresponding rows usually have different diameters (sensitivities) in the matrix, which does not affect differential privacy preservation on each PLS indeed. Finally, the true position is not input to produce a pseudo-location before the generation of the distribution matrix.
Such a procedure theoretically guarantees the privacy of the true location. The framework of DPIVE is shown in Fig. \ref{fig:DPIVE_model}.


DPIVE is mainly composed of two components: the partitioning algorithm $\mathcal{F}$ to determine disjoint PLSs and the differential privacy mechanism $\mathcal{K}$ to generate a pseudo-location. $\mathcal{F}$ has four inputs, prior distribution $\pi$, inference error threshold $E_m$, privacy parameter $\epsilon$ and location sets $\mathcal{X}=\{x_i\}$. For the two privacy parameters specified by users, $\epsilon$ allows users to control the posterior information leakage via the provisioning of differential privacy and $E_m$ aims to locally bound the expected inference error in the worst case.
Each PLS contains obviously at least two locations and ensures the lower bound of inference error.

Obviously, the result of our Algorithm $\mathcal{F}$ does not depend on the true location due to its no input. For minimization of the quality loss, $\mathcal{F}$ globally partitions the entire location domain into (as many as possible) disjoint PLSs satisfying \eqref{eq:inequality}.
Then, the mechanism $\mathcal{K}$ uses the diameter of each PLS, as the sensitivity of the exponential mechanism in corresponding $x_i$'s rows to calculate the probability distribution $f=\{f (x_j| x_i)\}$.
Afterwards, with the input of user's true location, DPIVE produces a pseudo-location via the public matrix $f$.

We mention that given the prior probability $\pi$ and the parameters $\epsilon$ and $E_m$, the PLSs partitioned in the dataset are determined by Algorithm \ref{alg:PLS}, and then the public matrix $f$ is computed and fixed. Moreover, the true location is  $\epsilon_g$-geo-indistinguishable among the locations within PLS, even in the worst case that the adversary knows the PLS. That is, DPIVE can provide users with location privacy protection satisfying their privacy requirements on $\epsilon$ and $E_m$ while the prior distribution $\pi$, Algorithm $\mathcal{F}$, differential privacy mechanism $\mathcal{K}$ and obfuscation probability matrix $\{f (x_j| x_i)\}$ are all public to the adversary. Besides, while in DPIVE any user has to employ unified privacy parameters of $\epsilon$ and $E_m$ for all regions, and in Section 4.6 we will consider the personalization of privacy budget. In the next two subsections, we present the details of Algorithm $\mathcal{F}$ and differential privacy mechanism $\mathcal{K}$, respectively.

\subsection{Partitioning Protection Location Sets}\label{}
\begin{figure}[tb]
\centering
	\subfigure[$4\times4$]{\label{fig:hilbert44}
		\includegraphics[scale=0.6]{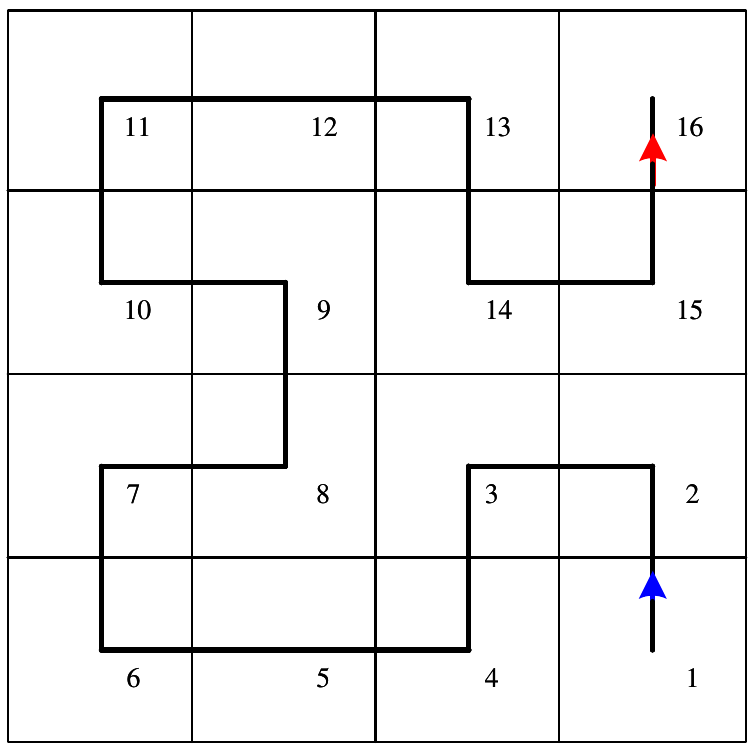}
	}\quad\quad\quad\quad\quad\quad
	\subfigure[$8\times8$]{\label{fig:hilbert88}
		\includegraphics[scale=0.6]{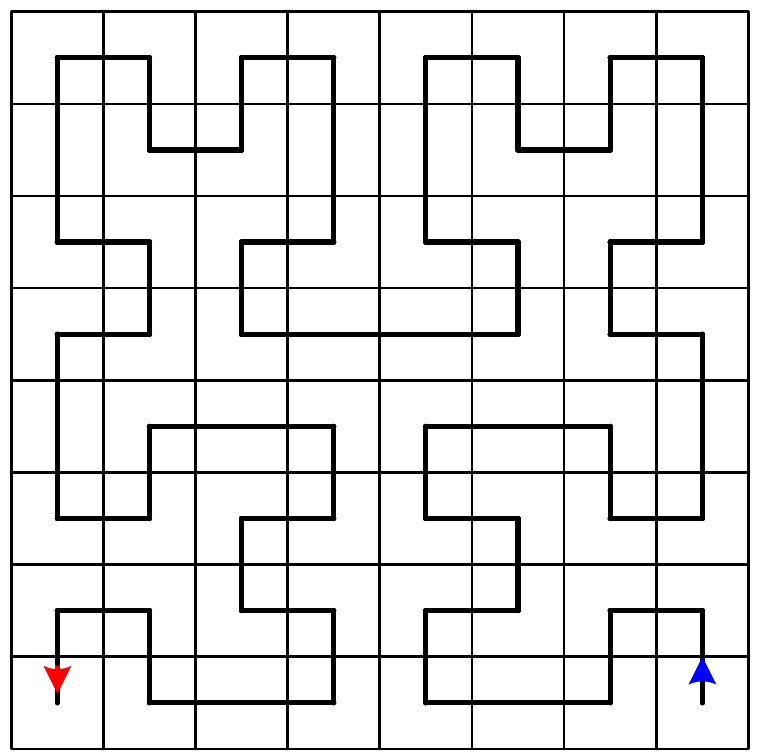}
	}
	\caption{Hilbert Curve for $4\times4$ and $8\times8$ grid.}
	\label{fig:hilbert}
\end{figure}

\begin{figure}[tb]
\centering
\subfigure[Locations along Hilbert Curve.]{\label{subfig:hilbert-black}
		\includegraphics[scale=0.5]{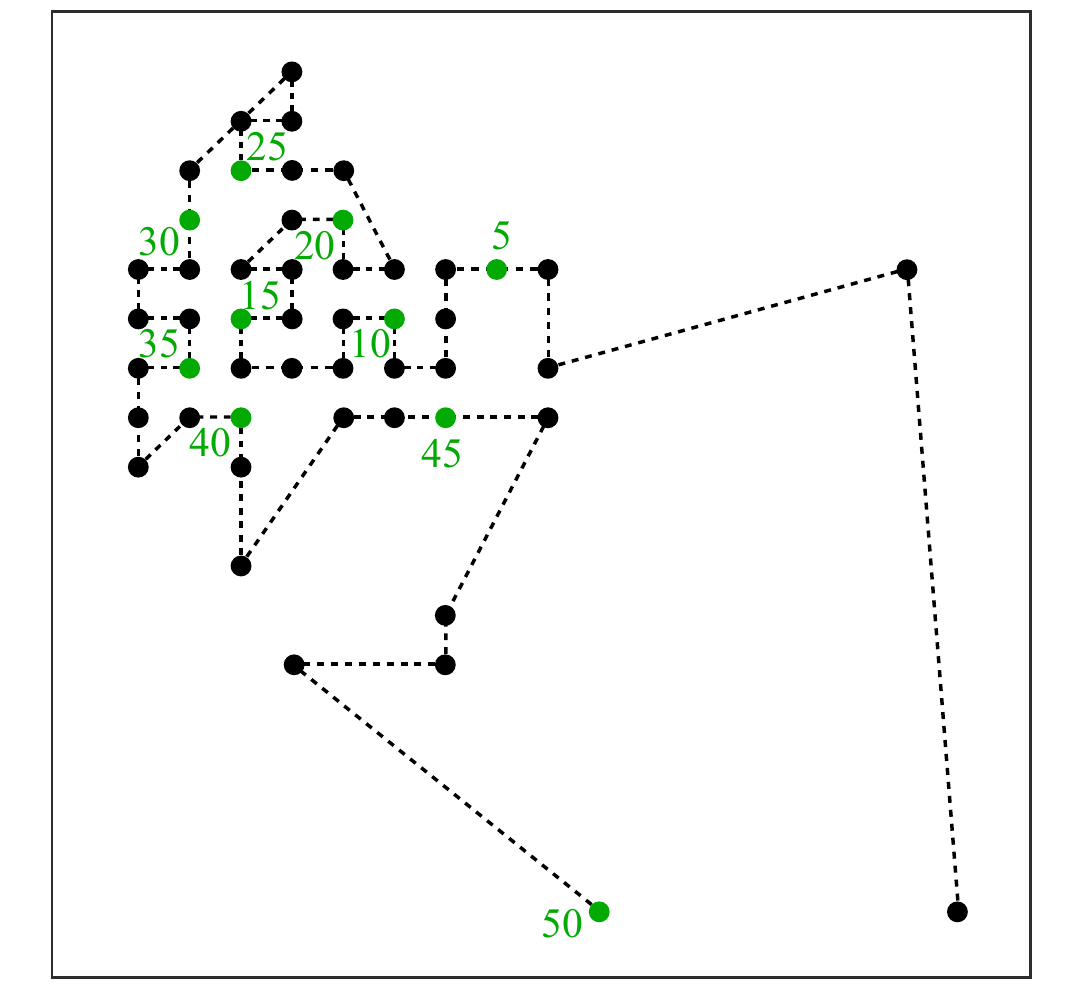}
	}\quad\quad\quad
\subfigure[Disjoint PLSs on Hilbert Curve.]{\label{subfig:hilbert-red-blue}
		\includegraphics[scale=0.5]{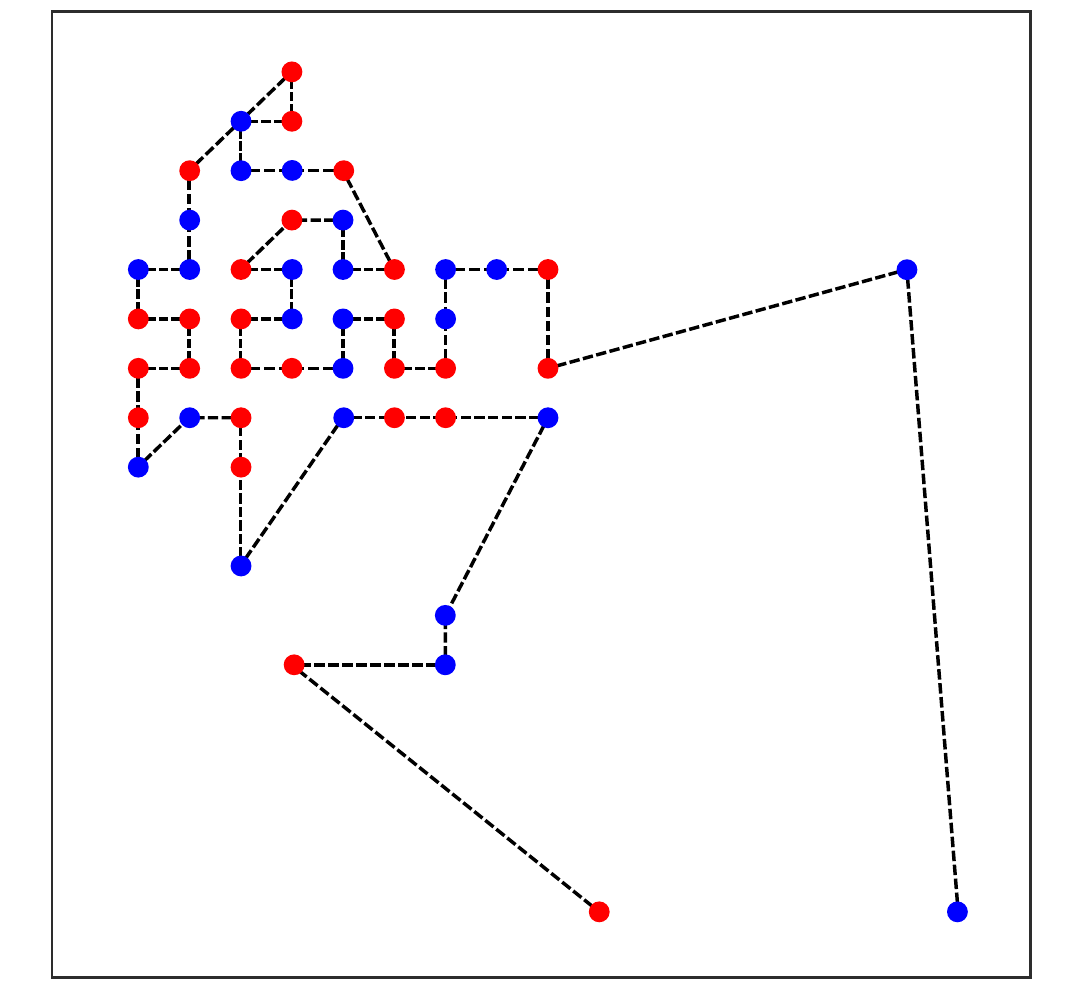}
	}
\caption{50 sequential regions on a Hilbert Curve.}
\label{fig:Geolife_hilbert}
\end{figure}

Hilbert curve \cite{LK00} is a common space-filling curve, which can map points in $2$-D space to one dimensional space and  has
the clustering properties with preserving the proximity of points. Fig. \ref{fig:hilbert} shows the Hilbert curves for $4\times 4$ and $8\times 8$ grids. Specifically, The curve maps a location point $x$ to a $1$-D value denoted by $H(x)$ called the Hilbert value of $x$, for example, Hilbert values $1$-$16$ of all cell centers in Fig. \ref{fig:hilbert44}. Following this, we connect the locations in the GeoLife dataset in the order of $H(x)$ and sort all locations in $\mathcal{X}$ with the rank denoted by $R(x)$, like $50$ points numbered in Fig. \ref{subfig:hilbert-black}.
It should be noted that the Hilbert curve generated in a $2$-D space is not unique. Rotating one Hilbert curve $90,\ 180,\ 270$ degrees clockwise around the center can generate other three Hilbert curves. For our regionalized location obfuscation mechanism, a region partition can only be performed on one Hilbert curve. In order to improve the performance of our mechanism, we execute Algorithm \ref{alg:PLS} independently on multiple (four) rotated Hilbert curves to perform region partitions and then choose the result with the smallest average diameter.

\begin{algorithm}[tb]
	\caption{Partitioning Algorithm for disjoint PLSs}
	\label{alg:PLS}
	\begin{algorithmic}[1]
		\REQUIRE
		sorted user's locations $\mathcal{X}=\{x_0,x_1,...,x_{n-1}\}$, prior probability $\pi$, inference error bound $E_m$, user privacy parameter $\epsilon$\\
		\STATE Initialize $\Phi_L=\{x_0,x_1\}$, $\Phi_R=\{x_{n-2},x_{n-1}\}$, $Q=\{x_2,\ldots,x_{n-3}\}$
		\STATE Remove ${x_i}^\prime s$ with the smallest subscript in $Q$ to $\Phi_L$ until satisfying \eqref{eq:inequality}
\label{state:start}
		\STATE Remove ${x_j}^\prime s$ with the largest subscript in $Q$ to $\Phi_R$ until satisfying \eqref{eq:inequality}
		\STATE \textbf{if} $|Q|\ge 2$ \textbf{then}
Add the set with the larger diameter between $\Phi_L$ and $\Phi_R$ to $\Phi_{\rm pls}$, initialize new $\Phi_L$ (if selected above) using points with the smallest subscript from $Q$, or new $\Phi_R$ similarly, and go to Line
\ref{state:start}

     \STATE \textbf{if} $|Q|= 1$, \textbf{then} Remove the only element to the nearer $\Phi_L$ or $\Phi_R$

     \STATE \textbf{if}
     \eqref{eq:inequality} holds for $\Phi_L$ and $\Phi_R$, \textbf{then} Add both $\Phi_L$ and $\Phi_R$ to $\Phi_{\rm pls}$
     and go to Line 10
		\STATE
\textbf{else} $\Phi_{RL} \gets \Phi_L\cup \Phi_R$
		\STATE \textbf{if} \eqref{eq:inequality} holds for $\Phi_{RL}$ \textbf{then} Add $\Phi_{RL}$ to $\Phi_{\rm pls}$
		\STATE \textbf{else} Bisect the curve $\Phi_{RL}$ and allocate the two parts
to two-sided neighbors from $\Phi_{\rm pls}$ with traversing for the smallest
average diameter in the sense of \eqref{small-diameter}
		\RETURN disjoint PLSs $\Phi_{\rm pls}$
	\end{algorithmic}
\end{algorithm}

Since we partition regions from a global perspective, the search range used in \cite{YLP17} can be omitted in Algorithm \ref{alg:PLS}.
Given a location set sorted according to the Hilbert curve, protection regions are constructed from the two sides of the curve to the middle and the initialized two alternatives are at the two ends, $\Phi_L$ and $\Phi_R$, respectively (Line 1).
Then supply $\Phi_L$ with neighboring locations on the right side along the curve one by one (Line 2) until that $\Phi_L$ is qualified for the condition \eqref{eq:inequality} and similarly supply for $\Phi_R$ (Line 3).

  If both $\Phi_L$ and $\Phi_R$ satisfy \eqref{eq:inequality}, assign the set with the larger diameter between $\Phi_L$ and $\Phi_R$ as a PLS to be removed into $\Phi_{\rm pls}$ (Line 4, isolated locations would be relatively preferred) and initialize new $\Phi_L$ or  $\Phi_R$ if removed. Process the steps by iterations (Lines 2-4) until $|Q|\le 1$, and afterwards we have to combine the remainder elements (Lines 5-8). If $\Phi_{RL}$ can not satisfy \eqref{eq:inequality} (Line 9), remove the locations with continuous rankings in $\Phi_{RL}$  to the two-sided $\Phi_j$'s on the Hilbert curve, and keep the new protection region satisfying \eqref{eq:inequality} and with the smallest diameter in the average sense of

 \begin{equation}\label{small-diameter}
 \frac{\pi(\Phi_1)\cdot D(\Phi_1) +\pi(\Phi_2)\cdot D(\Phi_2)}{\pi(\Phi_1)+\pi(\Phi_2)}, \ \ \ \text{with}\ \ \pi(\Phi_i)=\sum\limits_{x\in\Phi_i}\pi(x),\ \ i=1,2.
\end{equation}

There exists a situation with low probability, that is, no matter how the locations in $\Phi_{RL}$ are split for being allocated to its adjacent PLSs on two sides, \eqref{eq:inequality} is not satisfied for both new sets. Then the last (neighboring) set added to $\Phi_{\rm pls}$ can be combined with $\Phi_{RL}$, we assign the combination as new $\Phi_{RL}$ and return to Line 8 by iteration.

 The final disjoint PLSs partitioned by Alg. \ref{alg:PLS} is demonstrated by Fig. \ref{subfig:hilbert-red-blue}. Along the Hilbert curve, the neighboring locations marked in the same color (red or blue) belong to the same PLS. We mention that each PLS only includes some locations. Assuming that each location stands for a rectangle, the PLS is usually not a continuous region, since the covered locations are not neighboring on $H(x)$ in general.

\subsection{Exponential Mechanisms with Regionalized Sensitivity}\label{}

Given disjoint PLSs $\{\Phi_j\}$, DPIVE realizes differential privacy on each PLS $\Phi_j$ via the exponential mechanism \cite{DR14}. The set $\mathcal{X}$ is regarded as both input and output range of DPIVE. Since smaller distance produces higher utility, the utility of output location $x'$ can be measured by the Euclidean distance between perturbed and true locations, $d(x,x')$. The sensitivity of $u$ for each PLS $\Phi_j$ is
\begin{equation}\label{eq:sensitivity}
\Delta u(\Phi_j) = \max\limits_{x'\in \mathcal{X}}\max\limits_{x,y\in \Phi_j} |d(x,x')-d(y,x')|.
\end{equation}
Then from triangle inequality, we have $\Delta u(\Phi_j) = D(\Phi_j)$, i.e., the diameter of $\Phi_j$.

Since the disjoint $\Phi_j$'s are determined by the given privacy parameters instead of the true location, then
each input location (as true location) can not determine simply the sensitivity of $u$  and all locations in the same PLS $\Phi_j$ share the same sensitivity $D(\Phi_j)$.

{\bf Exponential Mechanism $\mathcal{K}$:}
 Given the disjoint sets $\{\Phi_j\}$ determined by privacy parameters $\epsilon$ and $E_m$ with satisfying \eqref{eq:inequality},
  for each apriori location $x\in\mathcal{X}$ and its corresponding PLS $\Phi_j$ derived from the given family $\{\Phi_j\}$,
the mechanism $\mathcal{K}$ computes
the probability distribution $f(x'| x)= w_{\Phi_j}(x) \exp\left(\frac{-\epsilon d(x,x')}{2D(\Phi_j)}\right)$ for any possible pseudo-location $x'$, where
\begin{equation}\label{wx}
w_{\Phi_j}(x) = \left( \sum_{x^\prime \in \mathcal{X} } \exp\left(\frac{-\epsilon d(x,x^\prime)}{2D(\Phi_j)}\right) \right)^{-1}.
\end{equation}

Following the public matrix $\{f (x_j| x_i)\}$, DPIVE mechanism generates a pseudo-location $x'\in\mathcal{X}$, which deploys user's true location information (to be protected with differential privacy) for the first time in the whole procedure.

We achieve $\epsilon$-differential privacy on each PLS and weak differential privacy on the whole domain as follows.

\begin{thm}[\cite{ZDC21}]\label{thm:DPIVE-DP}
Assume disjoint PLSs $\{\Phi_j\}$, then the exponential mechanism $\mathcal{K}$ in DPIVE satisfies $\epsilon$-differential privacy and
$(\epsilon_g,D(\Phi))$-geo-indistinguishability within each PLS $\Phi$.
\end{thm}



To be general, for the privacy preservation on whole $\mathcal{X}$, we have a weak assertion.

\begin{thm}[\cite{ZDC21}]\label{thm:region-symm}
Assume disjoint PLSs,  $\Phi_i$ and $\Phi_j (i\neq j)$, in the domain $\mathcal{X}$,  then
the exponential mechanism $\mathcal{K}$ in DPIVE satisfies $\left(\frac{D(\mathcal{X})}{D(\Phi_ i)}+\frac{D(\mathcal{X})}{D(\Phi_j)}\right)\frac{\epsilon}{2}$-differential privacy on $\Phi_i\cup\Phi_j$ and shortly $(\epsilon D(\mathcal{X})/D_{\min})$-DP on the whole domain $\mathcal{X}$, where $D_{\min}=\min_k D(\Phi_k)$.


\end{thm}

Theorem \ref{thm:region-symm} shows that any two locations from different PLSs are protected with weaker differential privacy.
This gives us a relatively complete result on the differential privacy preservation for the whole $\mathcal{X}$ no matter whether the two apriori locations are in the same PLS.

\subsection{Review of PIVE Framework}\label{subsect:PIVE}

In this subsection, we mainly recall the privacy problem of PIVE framework proposed in Yu et al. \cite{YLP17}, which is analyzed in detail in our previous work \cite{ZDC21}. Since our current DPIVE framework is a constructive correction of PIVE under the same assumption on parameter setting and Bayesian adversary model, it is enough for us to recall firstly their differences on the procedure and the privacy problem of PIVE. Indeed,
PIVE also includes two phases, as follows.

{\bf Phase I: Determining Protection Location Set}. The PLS for each location is generated adaptively and optimally.
PIVE regards $\Phi$ as a variable and dynamically searches region $\Phi$ satisfying \eqref{eq:inequality} with diameter as small as possible.

To be specific, for each input location $x$ denoted by $x_0$, the search algorithm returns a set having the smallest diameter satisfying \eqref{eq:inequality}.
 The locations in the output set are with consecutive rankings in $\mathcal{X}$ with respect to their mappings on a Hilbert curve.
Then each (true) location $x$ has its own PLS $\Phi_x$ and diameter $D(\Phi_x)$, and different (even neighboring) locations have different PLSs with different diameters. Even PLSs intersect with each other.

{\bf Phase II: Differentially Private Mechanism.}
The exponential mechanism is devised as above to generate pseudo-locations, which is desired (but failed) to
achieve differential privacy on the PLS. This is mainly due to the fact that different locations in the same PLS may have different diameters for applying the exponential mechanism.

For each PLS $\Phi_t$  determined by a true location $t$ and any $x,y \in \Phi_t$, we know in PIVE that $x$ and $y$ have their own PLS $\Phi_x$ and $\Phi_y$, respectively, and in general they have different sensitivities, i.e., the diameters $D(\Phi_x)\neq D(\Phi_y)$. Further, in the initial proof of differential privacy,

\begin{equation}\label{f-over-f-P}
\frac{f(x^\prime|x)}{f(x^\prime|y)}=
\frac{w_x \exp\left( -\epsilon d(x,x^\prime)/\left( 2D(\Phi_x) \right) \right)}{w_y \exp\left( -\epsilon d(y,x^\prime)/\left( 2D(\Phi_y) \right) \right)},
\end{equation}
we can not use the
 triangular inequality, $|d(x,x^\prime)-d(y,x^\prime)|\le d(x,y)$, in \eqref{f-over-f-P} as before. Thus, PIVE fails to achieve the guarantee of differential privacy as desired.

 Besides, the assumption narrowing adversary's guesses \emph{unfairly} to the private $\Phi$ fails to give the condition \eqref{ndss31-pri} for guaranteeing the minimum inference error $E_m$.
The corrected condition \eqref{eq:inequality} is shown by our Theorem \ref{thm:DPIVE} together with a counterexample, cf. Fig. \ref{fig:convex}. In conclusion, the main mistake of PIVE is derived from the adaptive search of PLSs.


\subsection{Region Partitioning by QK-means Clustering}\label{sec:qkmeans}
In this section, we partition the region back in the $2$-D space to achieve a more efficient privacy mechanism. Although the Hilbert curve method can well represent the proximity of locations in $2$-D space, it can only search the adjacent locations on the curve along a single direction, while the adjacent locations in $2$-D space may be far away from each other on the Hilbert curve (e.g., locations $2$ and $15$ in Fig. \ref{fig:hilbert44}).
 Even multiple Hilbert curves can not significantly improve the performance of the scheme. To overcome the limitations of the selection space on Hilbert curves, we design quasi $k$-means clustering (QK-means) algorithm via the popular $k$-means algorithm in machine learning. Basically we focus on constructing the Protection Location Set (PLS) including the true location and satisfying \eqref{eq:inequality}. When adding adjacent locations to the cluster, the QK-means method in $2$-D space has much more selections in clustering, unlike the Hilbert curve method in 1-D space.

Moreover, it is expected to achieve a suitable tradeoff between privacy protection and quality loss.  Some PLSs may be composed of only two locations for small privacy knobs, which will inevitably leak location privacy in the worst case that the adversary narrows the guesses within the PLS.
 For this, we can make a restriction on the smallest number of locations covered in every PLS, which is assigned as $2$ currently. Then we construct a partition for disjoint PLSs as many as possible with small diameter in the average sense.

\begin{algorithm}[h]
	\caption{Quasi K-means Clustering Algorithm}
	\label{alg:QK-means}
	\begin{algorithmic}[1]
		\REQUIRE sorted user's locations $\mathcal{X}=\{x_0,x_1,...,x_{n-1}\}$, prior probability $\pi$, inference error bound $E_m$, user privacy parameter $\epsilon$, maximum cyclic sampling times $Max\_Samp$, maximum iteration times $Max\_Iter$

		\STATE Init the number of clusters $k=1$ and $\tilde{\Phi}_1=\mathcal{X}$
		\STATE \textbf{if} \eqref{eq:inequality} is satisfied \textbf{then} $k=k+1$; \textbf{else return} $null$
		\STATE Init partitioned regions $\tilde{\Phi}_k=null$\label{state:begin}
		\STATE \textbf{Cyclic sampling:} Lines \ref{state:get_miu1}-\ref{state:cycle_end} for $Max\_Samp$ times
		\STATE Choose a loc from $\mathcal{X}$ randomly as the center $\mu_1$\label{state:get_miu1}
        \FOR {$j$ from 2 to $k$}
		\STATE Randomly choose a loc $x\in \mathcal{X}-\{\mu_1,\mu_2,...,\mu_{j-1}\}$ as $\mu_j$ with probability proportional to distance between $x$ and the set $\{\mu_1,\mu_2,...,\mu_{j-1}\}$
		\ENDFOR \label{state:get_other_miu}
        \STATE \textbf{Iteration:} Lines \ref{state:init_R}-\ref{state:cycle_end} for $Max\_Iter$ times
		\STATE Init remaining locations $Q=\mathcal{X}$, $\tilde{\Phi}=\{\Phi_1,\Phi_2,...,\Phi_k\}$ with $\Phi_j=\{\}(1\leq j\leq k)$\label{state:init_R}

		\STATE Remove each $x_i \in Q$ in ascending order of $\min_jd(x_i, \Phi_j)$ to its closest $\Phi_j$ that does not satisfy \eqref{eq:inequality} \label{state:get_dji}

		\STATE \textbf{if} \eqref{eq:inequality} is satisfied for all $\Phi_j$ in $\tilde{\Phi}$ \textbf{then} remove each remaining $x_i$ from $Q$ to its closest $\Phi_j$ with keeping satisfying \eqref{eq:inequality}

		\STATE \textbf{for} $j$ from 1 to $k$ \textbf{do} update $\mu_j=\frac{1}{|\Phi_j|}\sum_{x\in \Phi_j}x$
		\IF { \eqref{eq:inequality} is satisfied for all $\Phi_j$ in $\tilde{\Phi}$}
		\STATE \textbf{if} $\Phi_k=null$ or the average diameter $D(\tilde{\Phi})=\sum_k \pi(\Phi_k)D(\Phi_k)< D(\tilde{\Phi}_k)$ \textbf{then} $\tilde{\Phi}_k=\tilde{\Phi}$
        \ENDIF\label{state:cycle_end}
		\STATE \textbf{if} $\Phi_k \neq null$ and $D(\tilde{\Phi}_k)\leq D(\tilde{\Phi}_{k-1})$ \textbf{then} $k=k+1$ and go to Line \ref{state:begin}
		\RETURN disjoint PLSs $\tilde{\Phi}_{pls}=\tilde{\Phi}_{k-1}$
	\end{algorithmic}
\end{algorithm}

The QK-means method determines the final disjoint parts by adaptively searching for the optimal number of clusters $k$ as shown in Algorithm \ref{alg:QK-means}. For each $k$, the clustering centers are initialized on Lines \ref{state:get_miu1}-\ref{state:get_other_miu}. The first center is randomly selected in $\mathcal{X}$, and each subsequent center depends adaptively on those selected ahead, with sampling probability proportional to distance between each remainder location and its nearest center. This means that the longer the distance, the larger probability to be the new center, to make centers relatively sparse. On selecting locations to join the cluster, we search for the location each time that has the minimum distance to the centers (Line 11). Once a cluster satisfies \eqref{eq:inequality}, close it temporarily. If all clusters are closed, the remaining locations are added directly to their nearest clusters in order (Line 12). Then, improve the center by the mean vector in each cluster and carry out the next iteration until the mean vectors varies within a small range or the upper iteration times $Max\_Iter$ is achieved (Line 9). To eliminate the randomness of cluster center selection, we repeat sampling $Max\_Samp$ times on each $k$  (Line 4), for finding efficient partitioning that results in compact PLSs (with minimum average diameter) and satisfying \eqref{eq:inequality}. Increasing $k$ continues to find the next family of disjoint PLSs $\tilde{\Phi}_{k+1}$. If $\tilde{\Phi}_{k+1}$ can not be found or its average diameter is larger than $\tilde{\Phi}_k$, then $\tilde{\Phi}_k$ gives the final PLSs as required.

\begin{figure}
\begin{minipage}[t]{0.66\linewidth}
\vspace{0.75cm}
\centering
\subfigure[Geolife]{
		\includegraphics[scale=0.33]{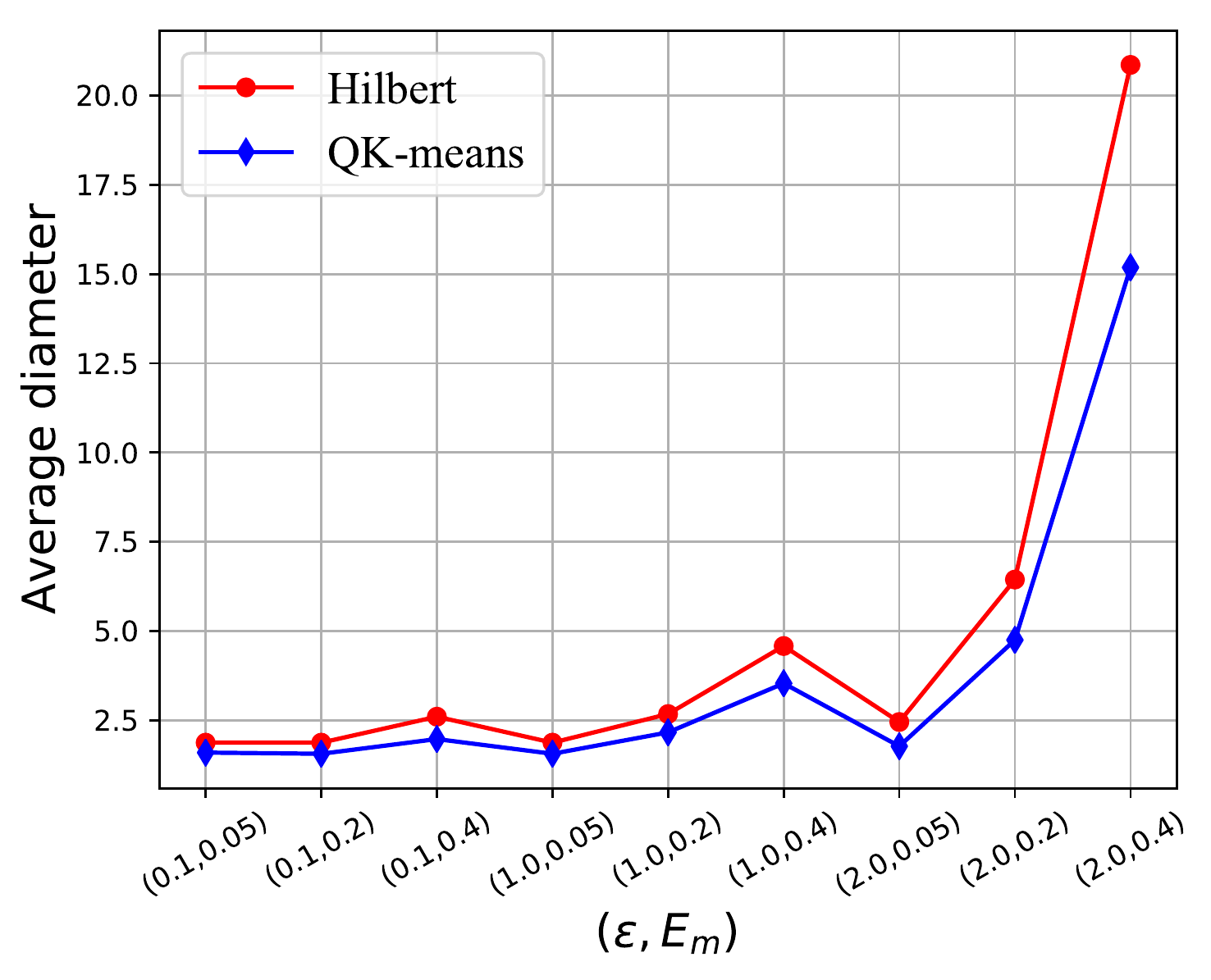}
	}
\subfigure[Gowalla]{
		\includegraphics[scale=0.33]{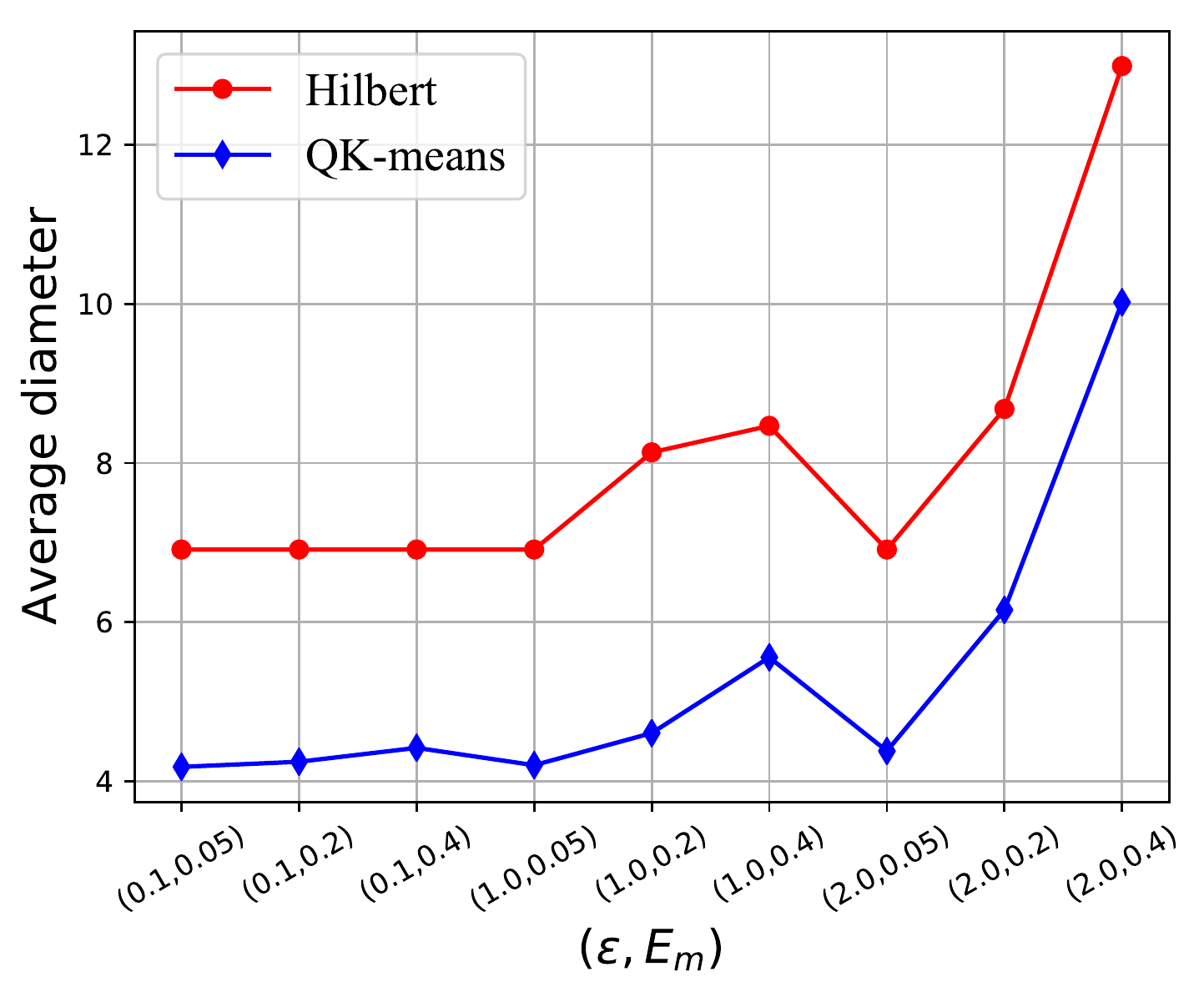}
	}
\caption{Effect of partition approaches for PLSs.}
\label{fig:diameter}
\end{minipage}
\begin{minipage}[t]{0.33\linewidth}
\vspace{0.01cm}
\centering
\includegraphics[height=2 in]{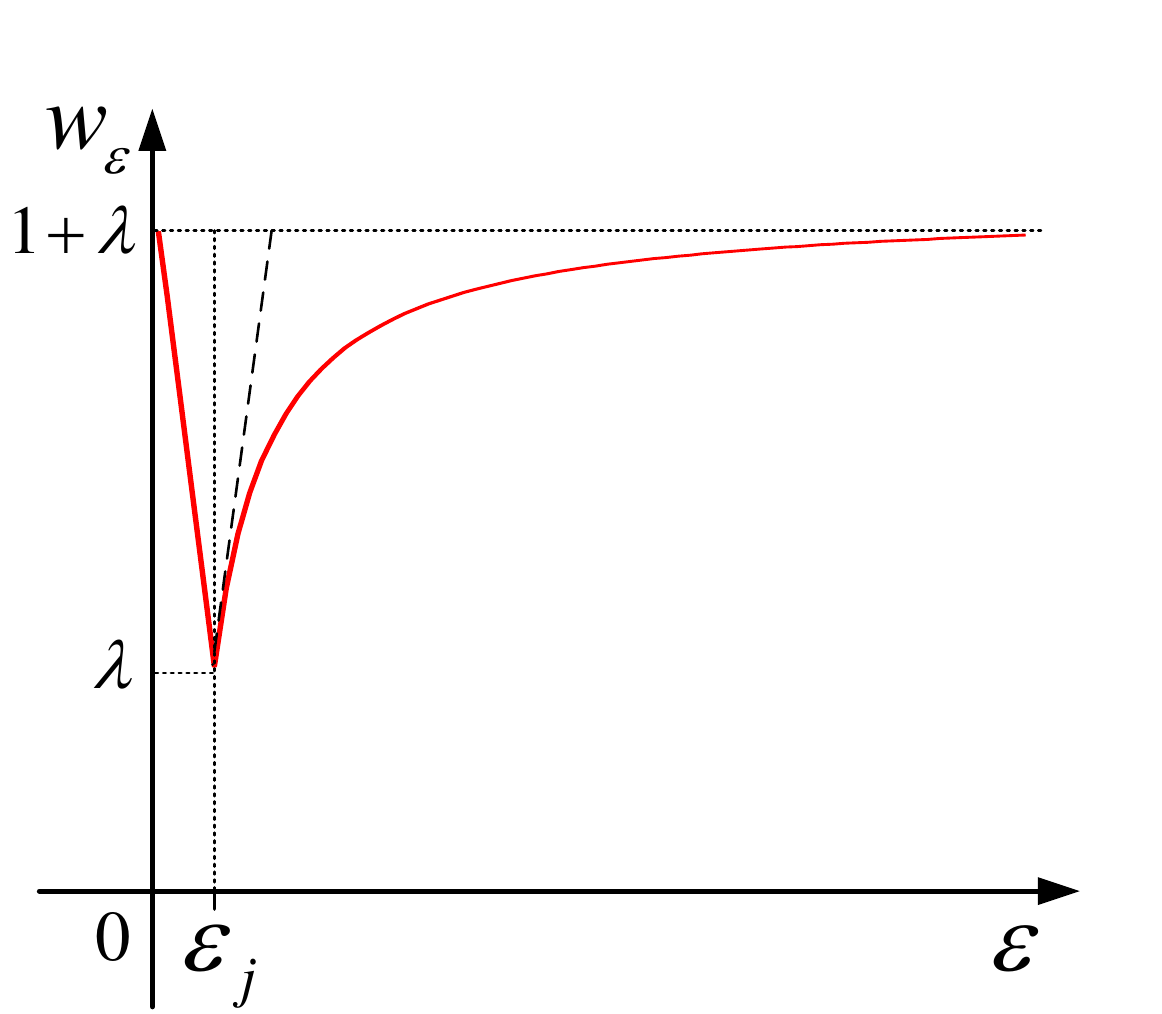}
\caption{Effect of varying $\epsilon$ on weight function for partitioning.}
\label{fig:function_pics}
\end{minipage}
\end{figure}

Fig. \ref{fig:diameter} compares the average diameter of the PLSs between Hilbert curve based method and QK-means method under different $\epsilon$ and $E_m$, in the sense of \eqref{small-diameter}. We sample three values of $\epsilon$ and $E_m$ separately to carry out $9$ groups of experiments on two datasets, respectively. The results show that on using QK-means, the globally average diameter is $21.8\%$ and 35.5\% smaller than that for Hilbert curve on GeoLife and Gowalla, respectively. More experiments will be executed in Section \ref{subsect:PDPIVE}.

\subsection{Personalizing \textbf{$\epsilon$}}\label{sec:personalization1}
Now we consider the personalization of user's privacy control knob. This allows users to set their privacy levels by customizing the privacy parameter $\epsilon$. The personalization of DPIVE mechanism is called PDPIVE.

 Different privacy levels of users generate different $\epsilon$, which brings some challenges to the search of PLSs. As we know, the PLSs constructed in DPIVE result in the same privacy level for users due to \eqref{eq:inequality}. In order to satisfy the privacy requirements of all locations within the same PLS $\Phi_j$, DPIVE has to achieve the user's highest privacy level in $\Phi_j$, that is, the region's privacy budget $\epsilon_j=\mathop{\min}\limits_{x\in \Phi_j} \epsilon_x$ due to \eqref{lower-bound-sum} theoretically.

To ensure the lower bound of expected inference error, we can obtain the claim as follows based on Theorem \ref{thm:DPIVE}.

\begin{thm}[]\label{thm:PDPIVE}
Given a domain partition $\{\Phi_k\}$ and an observed pseudo-location $x^\prime$ in $\mathcal{X}$, suppose that an obfuscation mechanism satisfies $\epsilon_k$-DP on each PLS $\Phi_k$. If $E'(\Phi_k)\geq e^{\epsilon_k}E_m$ for each $\Phi_k$, then
 $ExpEr(x^\prime)\ge E_m$ for the optimal inference attack.
  \begin{proof}
 Given that the obfuscation mechanism satisfies $\epsilon_k$-DP on each PLS $\Phi_k$ from a partition $\{\Phi_k\}$,
 we obtain, by normalization in each PLS $\Phi_k$,
\begin{equation}\label{lower-bound-sum-k}
ExpEr(x')\ge  
\sum_{k} \text{Pr}(\Phi_k|x') e^{-\epsilon_k}  E'(\Phi_k).
\end{equation}
\noindent

\noindent 
Since $\sum_{k} \text{Pr}(\Phi_k|x')=1$, 
the condition
that for all $\Phi_{k}$,
\begin{equation}\label{ndss31-pri-k}
E'(\Phi_k)\ge e^{\epsilon_k} E_m,
\end{equation}
implies the user-defined error threshold, $ExpEr(x^\prime)\ge E_m$, for the optimal inference attack using any observed pseudo-location $x'$.
 \end{proof}

\end{thm}

In this scenario, the privacy parameter $\epsilon$ has to be considered on partitioning the region. Adding each location to a PLS may affect the privacy level of PLS. However, current QK-means considers only the distance while ignoring the differences on $\epsilon$ among locations.  For this,
the Euclidean distance $d_{ji}$ between $x_i$ and $\Phi_j$ used on Line \ref{state:get_dji} of Algorithm \ref{alg:QK-means} is replaced by $\sigma_{ji}=d_{ji}\cdot w_\epsilon$ with weight $w_\epsilon$ emphasizing the influence of $\epsilon$ on $\sigma_{ji}$,
\begin{equation}\label{eq:w_ep}
w_\epsilon=1+\lambda-\frac{\min(\epsilon, \epsilon_j)}{\max(\epsilon, \epsilon_j)},
\end{equation}
where $\epsilon_j$ represents the current privacy budget of the PLS $\Phi_j$ that is to be updated once a new location with privacy $\epsilon$ is added, $\lambda$ is a parameter to control the range of $w_\epsilon$ and the default value of $\lambda$ is $0.5$. Such a setting prefers those locations with $\epsilon$ value more than and closed to current $\epsilon_j$, see Fig. \ref{fig:function_pics}(a). Indeed, the newly added location with smaller  $\epsilon$ will certainly modify the current $\epsilon_j$ which probably produces larger quality loss, while the added location with larger $\epsilon$ will not change the $\epsilon_j$. The parameter $\lambda$ aims mainly to avoid the case of $w_\epsilon=0$ that totally ignores the effect of distance.

\begin{figure}[htbp]
\centering
\subfigure[Geolife]{
		\includegraphics[scale=0.45]{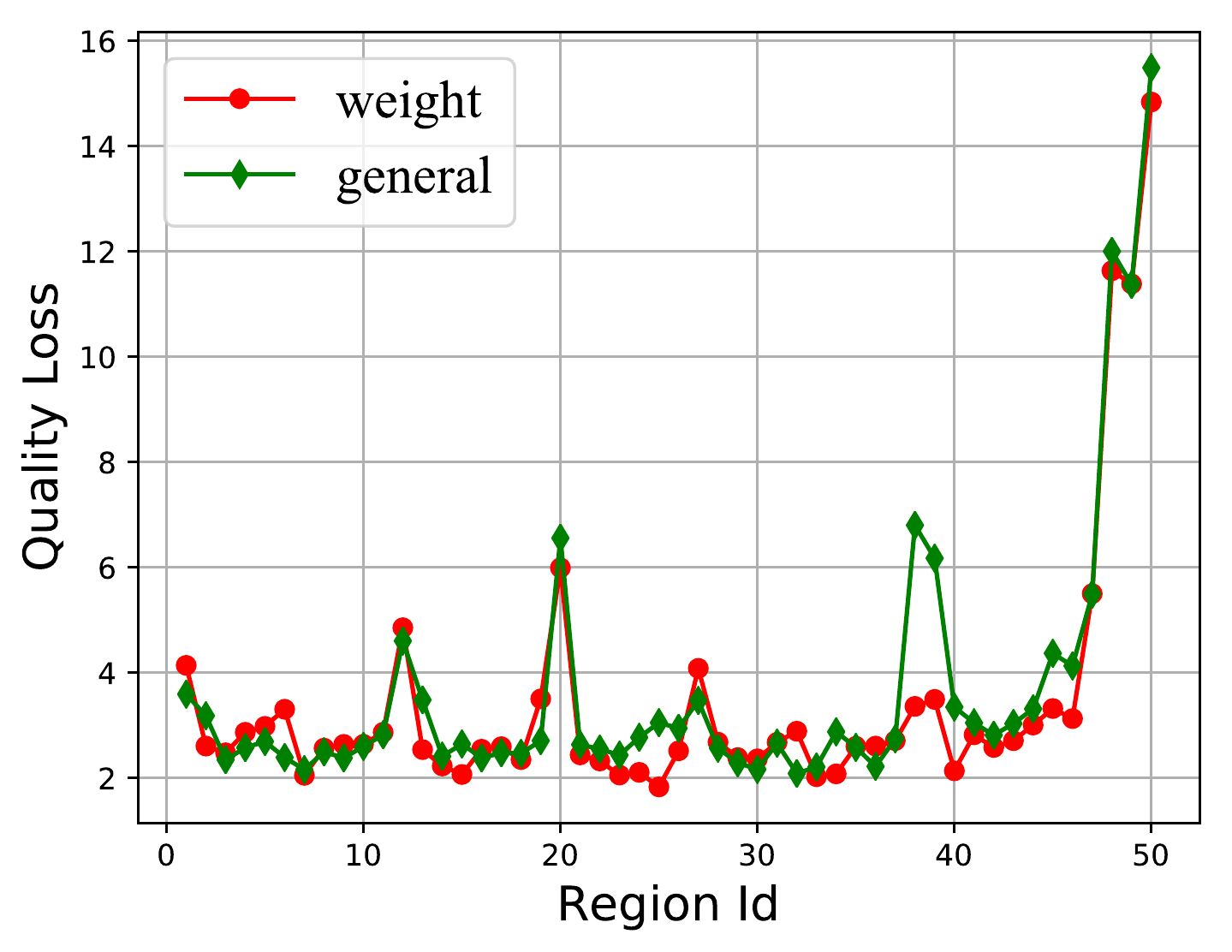}
	}
\subfigure[Gowalla]{
		\includegraphics[scale=0.45]{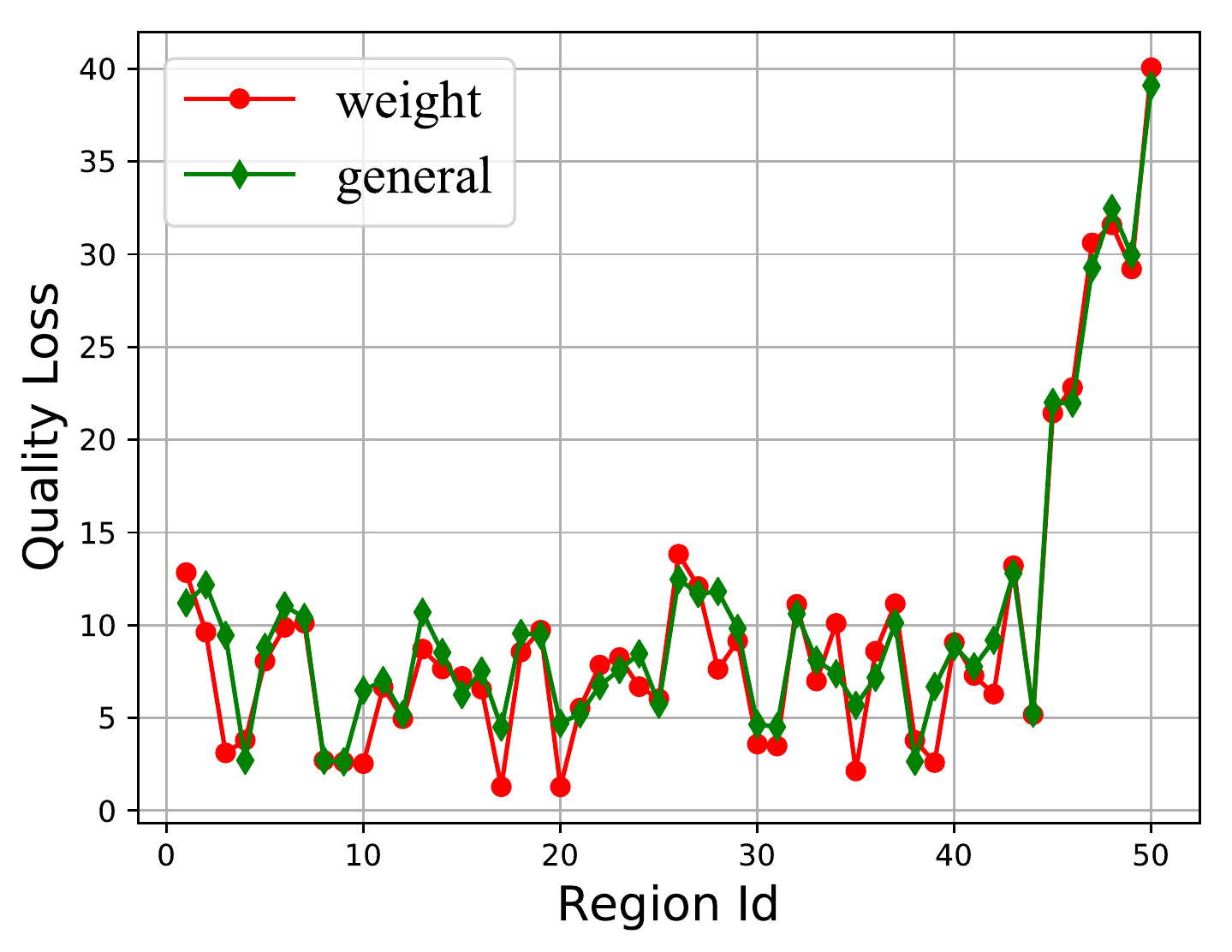}
	}
\caption{Effect of adding weights on PDPIVE.}
\label{fig:weight}
\end{figure}

We test the effect of $w_{\epsilon}$ on the GeoLife dataset.
Based on DPIVE which adopts QK-means algorithm, two strategies are adopted in the clustering process respectively, one is the $weight$ scheme using the above weight, and the other is the $general$ scheme without the use of weight ($w_{\epsilon}=1$).
The $\epsilon$ of each location is uniformly and randomly sampled in $[0.5,1.5]$ to simulate the $\epsilon$ of user personalization and $E_m=0.1$ is fixed. The quality loss of two schemes is shown in Fig \ref{fig:weight}.

The experimental results show that the average quality loss decreases from 3.69 to 3.44 in GeoLife and 10.33 to 9.7 in Gowalla by taking weights into accounts, respectively. The quality loss on half locations is obviously improved. This demonstrates that such weights make more locations with closer privacy levels on $\epsilon$ join in the same PLS, which effectively reduces the service quality loss.

Next, on the real-world location-based service applications, as mentioned in Fig. \ref{fig:DPIVE_model}, both control knobs, minimum inference error and differential privacy parameter, are assumed to be private for each user. Algorithm $\mathcal{F}$, differential privacy mechanism $\mathcal{K}$ and obfuscation probability matrix $\{f (x_j| x_i)\}$ are all public to adversaries, and they are used locally by the user to produce a pseudo-location. Each user can define their differential privacy parameter personally on each location in the following two provided ways: 1) detailed operation instruction with some prime examples; and
2) default setting for different privacy levels, like conservative (small value), moderate (middle value)
and liberal (great value) levels, in which the concrete knob values for each level can be adjusted appropriately.

\section{Performance Valuation}\label{sec:performance}

We first compare our
DPIVE approach with some previous
mechanisms on the metrics of location privacy and service quality, then
 present an experimental evaluation
of PDPIVE scheme.
The results show that our mechanisms effectively combine both privacy
notions and efficiently address privacy protection issues on isolated locations.

\subsection{Experimental Methodology}
\begin{figure}[tb]
\centering
\subfigure[Geolife]{
		\includegraphics[height=1.8 in]{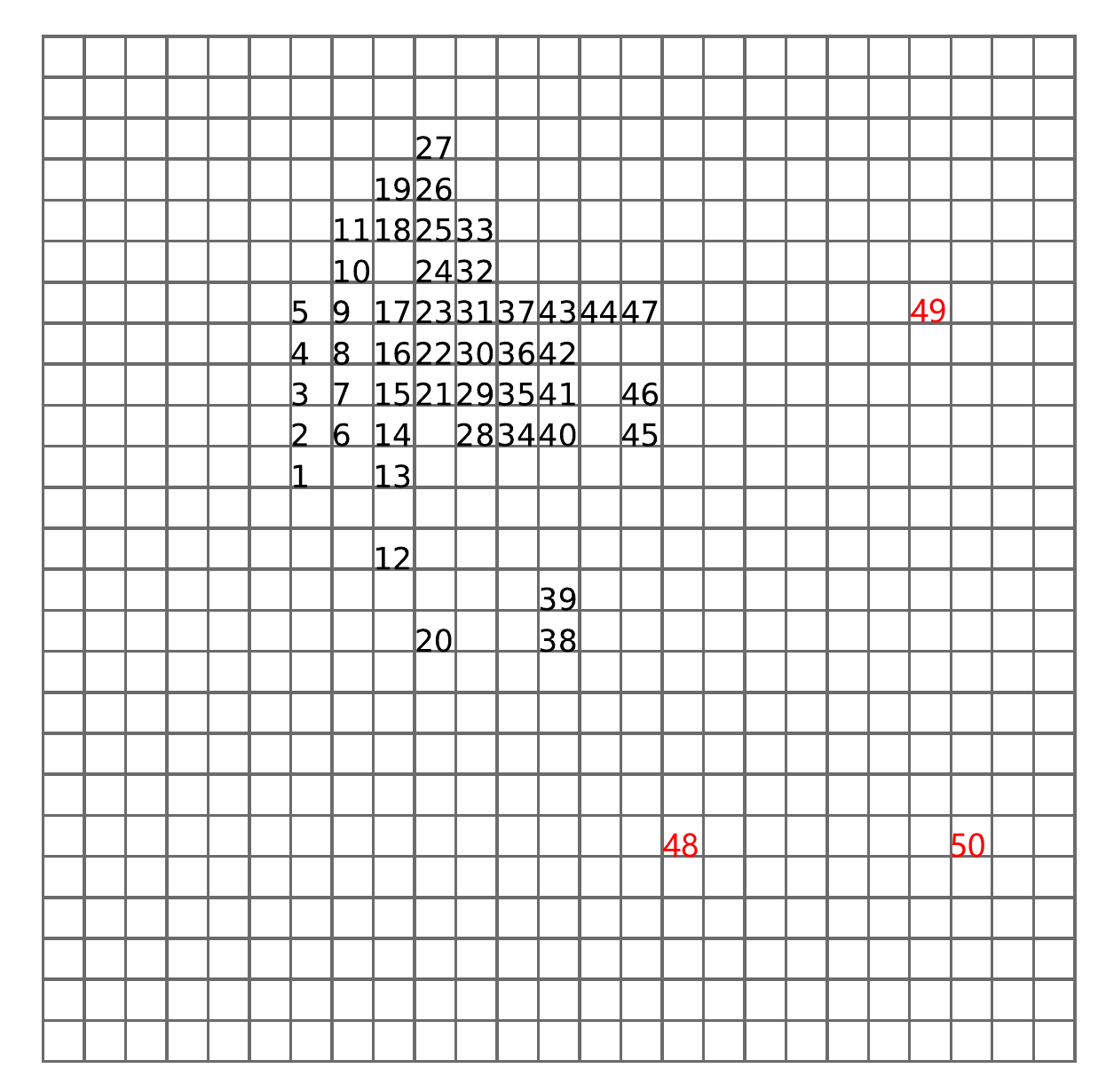}
	}\quad\quad\quad\quad\quad\quad
\subfigure[Gowalla]{
		\includegraphics[height=1.8 in]{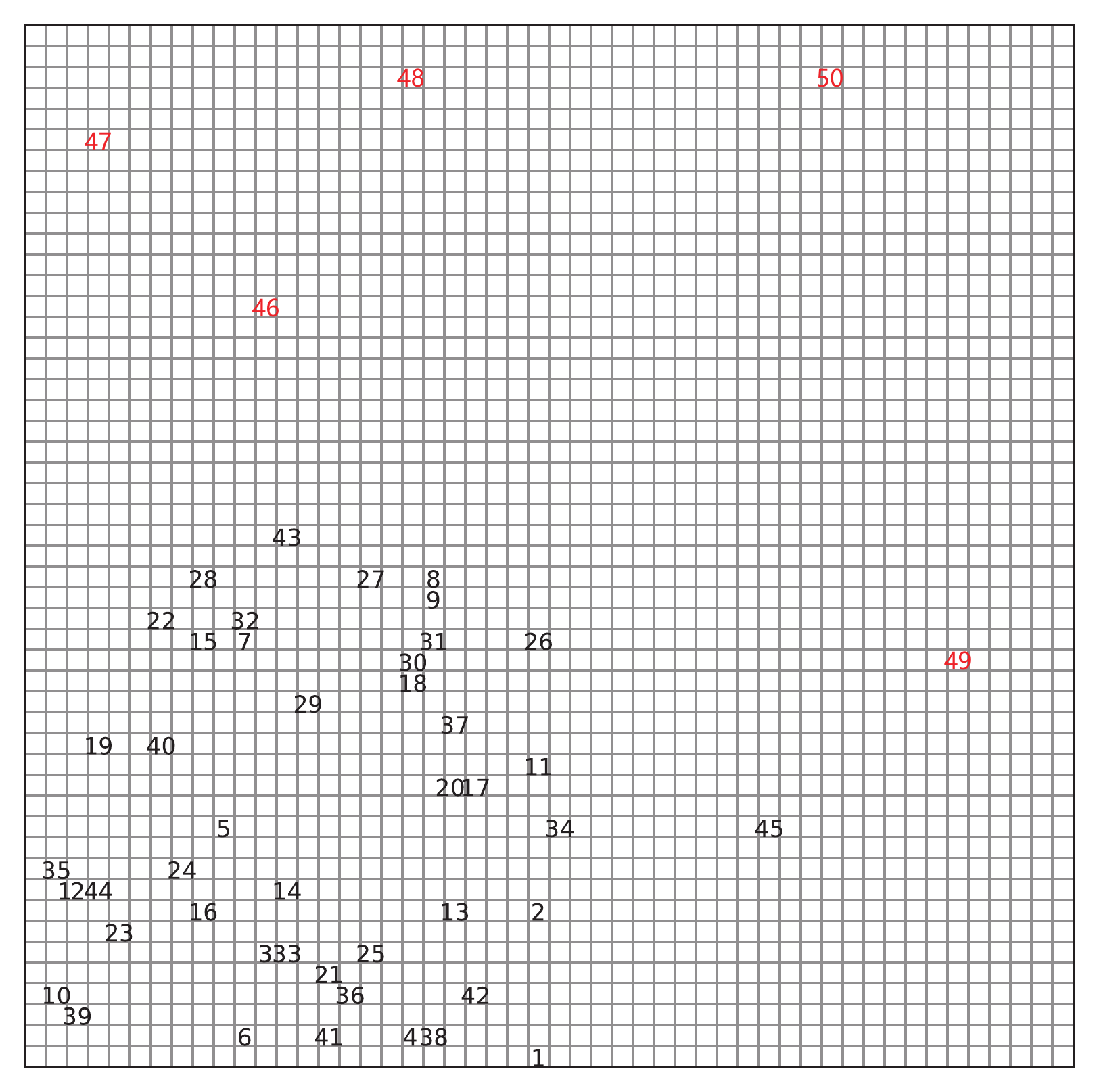}
	}
\caption{50 regions distributed in two datasets.}
\label{fig:datasets}
\end{figure}

\textbf{Datasets.} Two location sets are used in the experiment, which are extracted from GeoLife and Gowalla datasets, respectively. The location distribution in GeoLife is relatively dense, while sparse in Gowalla. For GeoLife, we use the same distribution as \cite{YLP17}, and for convenience we assign the grid size as $1$km$\times1$km. Gowalla is a social network check-in dataset containing $224$ days of check-in data for California in 2010. We divide the main area of Gowalla into also $1$km$\times1$km cells and make random selections for $50$ relatively sparse cells. The distributions of both datasets are shown in Fig. \ref{fig:datasets}, in which most isolated regions are numbered behind.

We simulate a prior distribution uniformly on both datasets, in which each value is sampled randomly and uniformly in $[0.01, 0.03]$ with normalization, see Table \ref{tlb:pi}.

\begin{table}[htbp]\renewcommand{\arraystretch}{1.5}
	\centering
    \small
	\caption{Values of prior probability ($\times10^{-2}$).}\label{tlb:pi}
	\begin{tabular}{|c|c|c|c|c|c|c|c|c|c|c|}
    \hline
    \textbf{1-10}&1.53&2.41&1.11&1.23&2.29&2.00&2.13&2.06&1.87&1.43\\\hline
    \textbf{11-20}&1.84&2.24&1.54&1.50&1.50&2.53&2.15&2.59&2.46&1.90\\\hline
    \textbf{21-30}&2.43&2.10&2.46&1.62&1.50&1.32&2.55&1.97&2.61&2.82\\\hline
    \textbf{31-40}&2.69&2.27&1.81&1.79&2.78&2.84&1.66&2.69&1.07&1.99\\\hline
    \textbf{41-50}&1.99&1.92&1.06&2.49&1.09&2.68&1.93&2.40&1.84&1.64\\\hline
	\end{tabular}
\end{table}

\textbf{Parameters setting.} The lower bound of inference error $E_m\in\{0.05,0.1,...,0.5\}$. The privacy budget $\epsilon\in\{0.1,0.3,\ldots,$ $1.9,2.0\}$ in GeoLife and $\epsilon\in\{0.1,0.3,\ldots,2.5\}$ in Gowalla. The reason for the difference on budget range is that large $\epsilon$ would imply large PLS for satisfying the condition \eqref{eq:inequality} and particularly the whole (relatively dense) 50-point dataset GeoLife can not satisfy \eqref{eq:inequality} as a PLS with $\epsilon=2.1$ for some $E_m$.

On the aspect of personalization, randomly and uniformly sampling parameters is restricted in the middle of the above ranges,  $\epsilon \in[0.5,1.5]$. In order to measure the performance improvement brought by personalized mechanism, we assign DPIVE scheme as baseline that uses unified privacy parameters for the whole $\mathcal{X}$. Specifically, in order to meet the highest privacy requirements of all PLSs, $\epsilon=0.5$ if personalized.

\subsection{Performance Analysis of DPIVE}

\textbf{Comparing the protection of skewed locations.} In this section, we compare DPIVE (using Hilbert curve based method) with previous typical mechanisms, EM \cite{YLP17}, Joint \cite{Sho15} and Opt-Geo \cite{BCP14}, especially to verify the advantages of DPIVE on protecting isolated regions as in \cite{YLP17}. Rather than the globally average performance of privacy protection emphasized in previous work, DPIVE pays more attention to the local performance. Then we also check the detailed privacy protection performance on each region. In order to make a fair comparison between different schemes, we specify the parameters of DPIVE ($\epsilon=1.0$, $E_m=0.05$) and adjust the parameters of other schemes to ensure the same location privacy, that is, the same unconditional expected inference error.

The EM mechanism is similar to the exponential mechanism proposed in PIVE, except that a constant diameter is used for the protection region of each location. EM adopts the same $\epsilon$ as DPIVE and adjusts the constant diameter ($1.66$km) so that their expected inference errors achieve the same (their difference within $0.005$ is acceptable).

Opt-Geo is an efficient privacy mechanism that minimizes quality loss through linear programming while satisfying geo-indistinguishability. We use $\delta=0.05$ commonly as in \cite{BCP14} and determine $\epsilon_g=0.3$ to reach the same expected inference error.

Joint is the first mechanism that uses linear programming to combine two privacy notions of expected inference error and geo-indistinguishability. We use the same $\epsilon=1.0$, and then use DPIVE's global expected inference error as the minimum desired distortion privacy level $d_m$, via adjusting $\epsilon_g=0.3$ to obtain the same expected inference error.

The scheme privacy is measured by the average inference error $AvgErr$ of the optimal inference attack and success probability $p_s$ of Bayesian inference attack \cite{YLP17}. Define

\begin{equation}
AvgErr(x)=\sum_{x^\prime \in \mathcal{X}}f(x^\prime | x)d(\hat{x},x),
\end{equation}
\begin{equation}
p_s(x)=\sum_{x^\prime \in \mathcal{X}}f(x^\prime | x)d_h(\hat{x},x),
\end{equation}
where $\hat{x}$ (determined by $x'$) is obtained by \eqref{eq:attack x1} for $AvgErr$ with $d(\hat{x},x)$ representing Euclidean distance while obtained by \eqref{eq:attack x2} for $p_s$ with $d_h(\hat{x},x)$ denoting Hamming distance.

\begin{figure}[tb]
\centering
\includegraphics[scale=0.25]{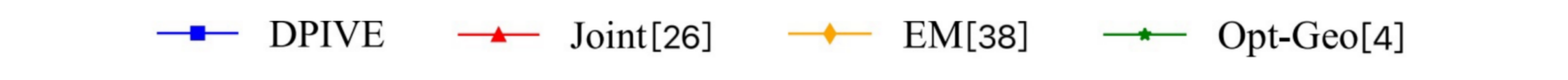}\\
\subfigure[Optimal inference attack (Geo.)]{
		\includegraphics[scale=0.3]{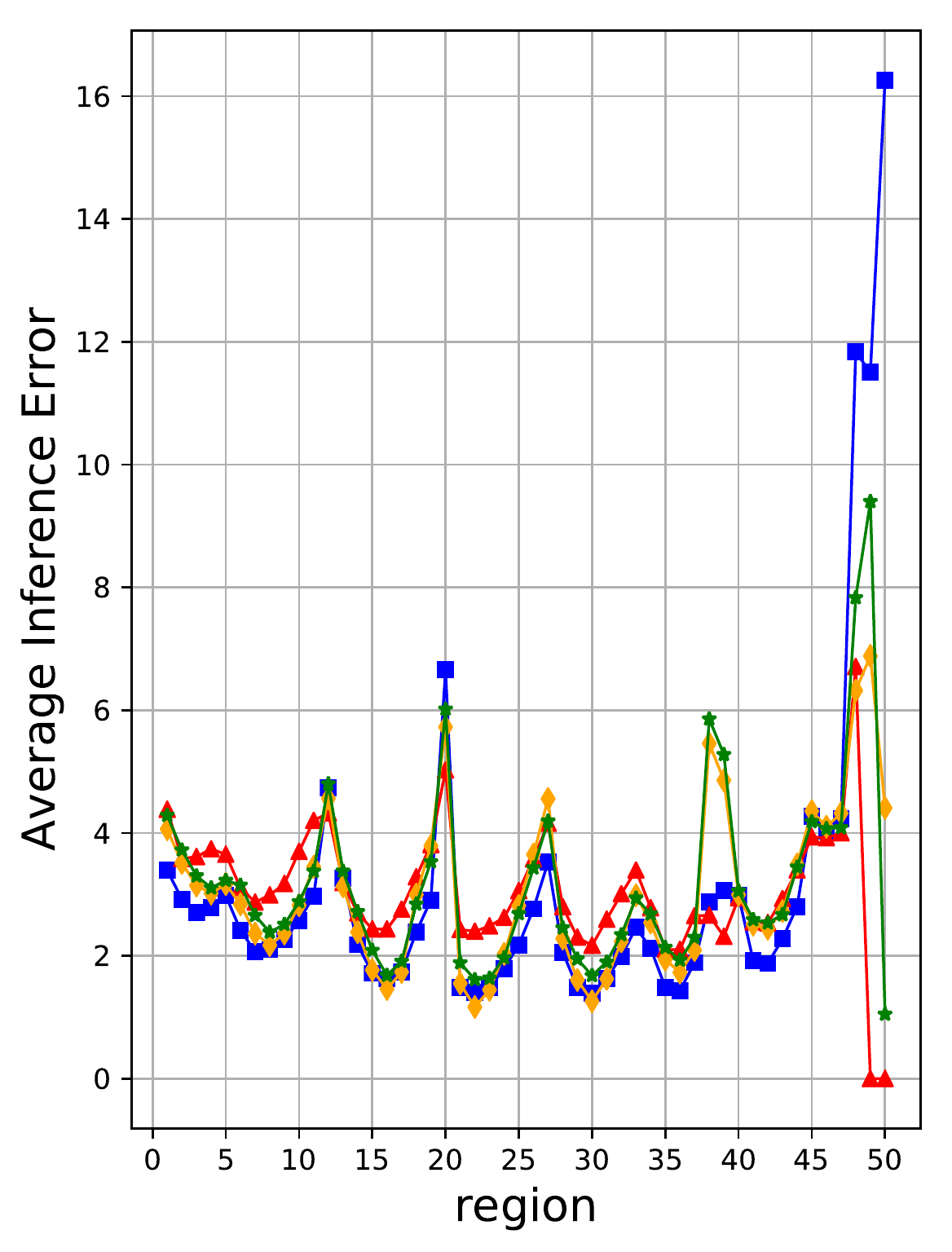}
	}
\subfigure[Bayesian inference attack (Geo.)]{
		\includegraphics[scale=0.3]{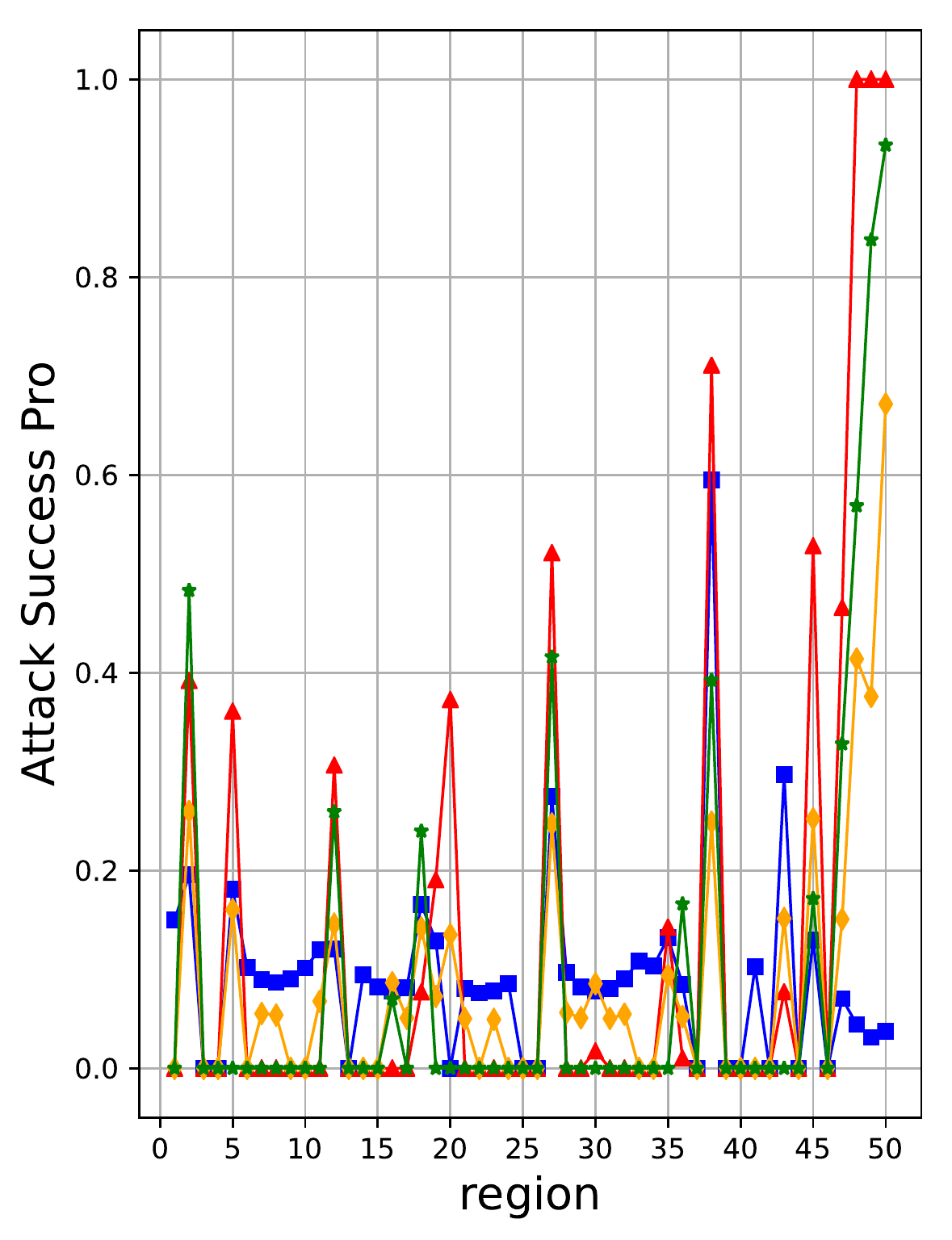}
	}
\subfigure[Optimal inference attack (Gow.)]{
		\includegraphics[scale=0.3]{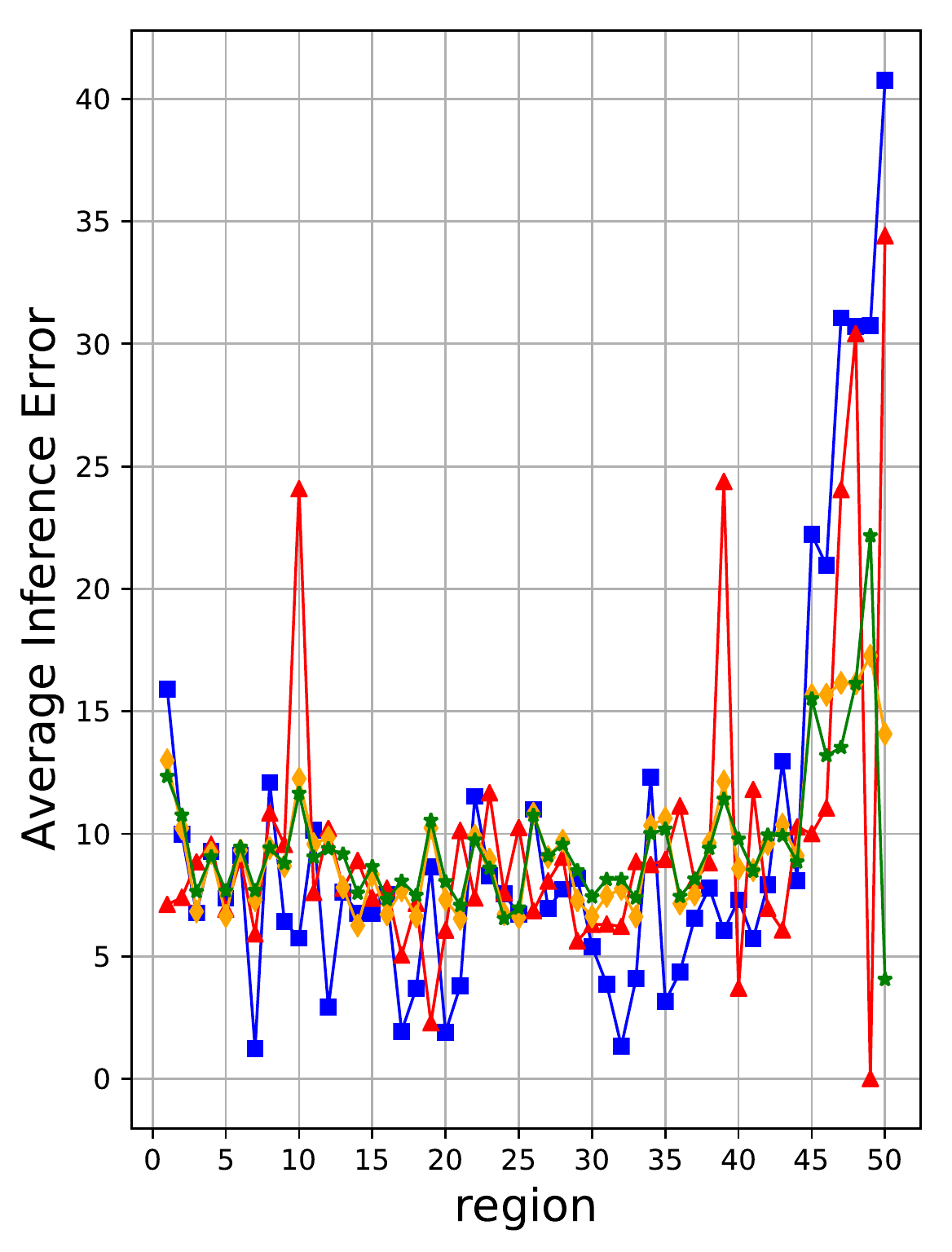}
	}
\subfigure[Bayesian inference attack (Gow.)]{
		\includegraphics[scale=0.3]{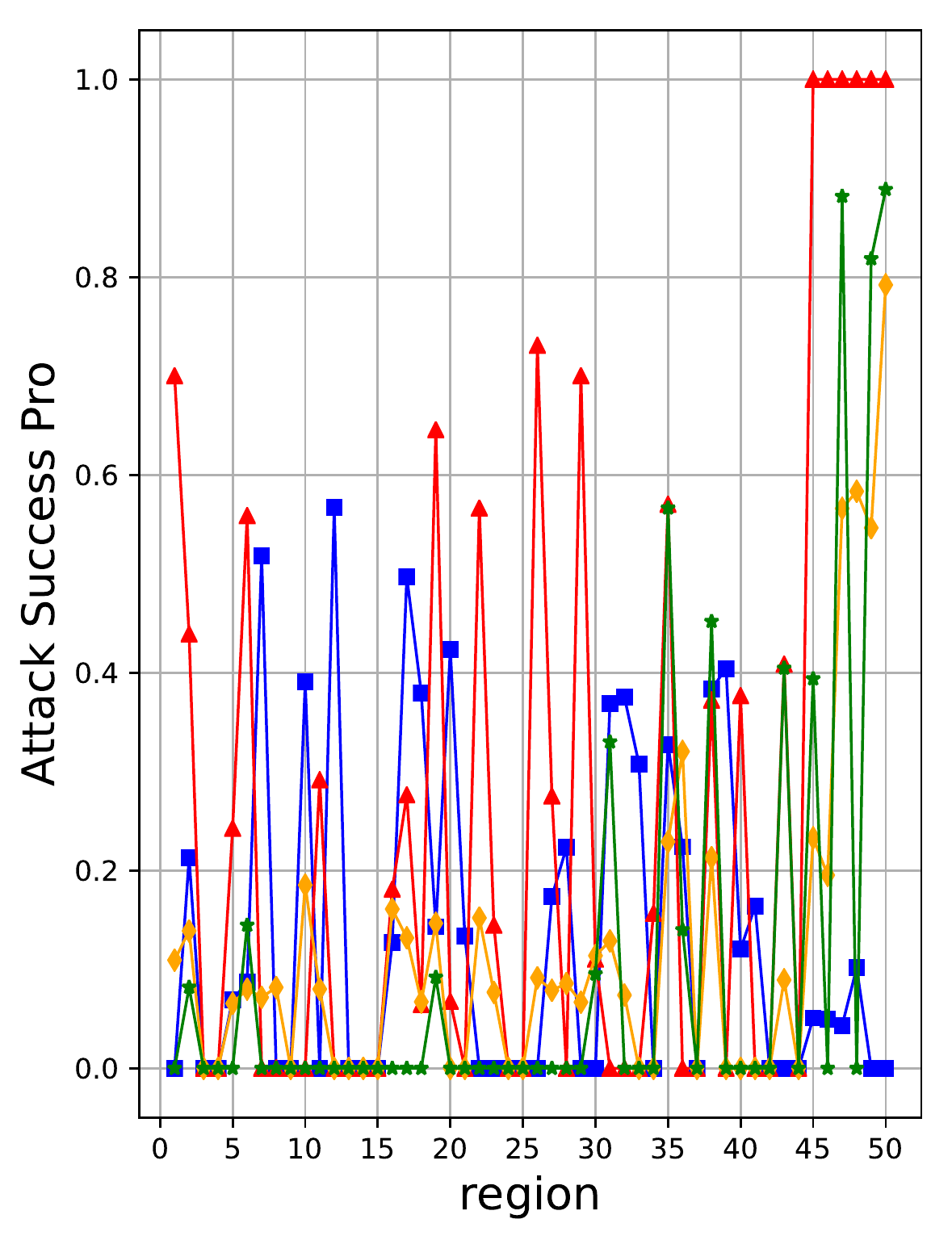}
	}
\caption{Comparison of DPIVE with Joint, EM, and Opt-Geo.}
\label{fig:DPIVE}
\end{figure}


Fig. \ref{fig:DPIVE} shows the comparisons of the average inference error and expected success probability of Bayesian inference attack (using Hamming distance) on each region among four mechanisms. Due to the above adjustments for reaching the same unconditional expected inference error for four schemes, DPIVE has a lower average inference error \emph{AvgErr} in most regions while it has higher \emph{AvgErr} on isolated regions than the other schemes. It does not mean that DPIVE is easier to be attacked, and the analysis is as follows.

In some isolated regions (such as $48$-$50$ in GeoLife and $46$-$50$ in Gowalla, marked in red in Fig. \ref{fig:datasets}), the schemes EM, Opt-Geo and Joint have a significant increase in the expected success probability, even the Joint reaches $100\%$ (accurate attack), while DPIVE has less than $20\%$. Indeed, DPIVE partitions the local protection region according to the privacy parameters $\epsilon$ and $E_m$ to ensure the lower bound of inference error in the worst case, thus it effectively and locally protects the isolated regions.

\begin{table}[tb]\renewcommand{\arraystretch}{1.5}
	\centering
    \small
	\caption{The percentage of locations exceeding each success probability threshold, and quality loss.}\label{tlb:attack_pro}
    \resizebox{\textwidth}{20mm}{
	\begin{tabular}{|c|c|c|c|c|c|c|c|c|c|}
    \hline
    \multicolumn{2}{|c|}{\textbf{Dataset}}&\multicolumn{4}{c|}{\textbf{GeoLife}}&\multicolumn{4}{c|}{\textbf{Gowalla}}\\\hline
    \multicolumn{2}{|c|}{\diagbox[width=9em]{\textbf{Metrics}}{\textbf{Schemes}}}&
    \textbf{DPIVE}&\textbf{EM \cite{YLP17}}&\textbf{Opt-Geo \cite{BCP14}}&\textbf{Joint \cite{Sho15}}&\textbf{DPIVE}&\textbf{EM \cite{YLP17}}&\textbf{Opt-Geo \cite{BCP14}}&\textbf{Joint \cite{Sho15}}\\
    \hline
    \multirow{3}*{\textbf{$X.$}}&\textbf{\textgreater50\%}&2\%&2\%&6\%&12\%&4\%&8\%&8\%&26\%\\\cline{2-10}
    &\textbf{\textgreater70\%}&0\%&0\%&4\%&8\%&0\%&2\%&6\%&18\%\\\cline{2-10}
    &\textbf{\textgreater90\%}&0\%&0\%&2\%&6\%&0\%&0\%&0\%&12\%\\\hline
    \multicolumn{2}{|c|}{\textbf{Quality Loss}}&3.22&3.27&3.12&3.9&9.88&9.93&9.46&9.98\\\hline

    \end{tabular}}
\end{table}

Moreover, under the premise of the same location privacy requirements, we count the percentage of regions whose attack success probability exceeds $X\%$ for each scheme as shown in Table \ref{tlb:attack_pro}. It demonstrates that DPIVE has always the lowest attack success probability when $X$ takes $50\%,\ 70\%$, and $90\%$, even there are no regions that have attack success rate higher than $60\%$.

%

On the aspect of quality loss, Opt-Geo achieves the smallest quality loss due to its global optimization on service quality, Joint has the highest, and DPIVE is close to EM. It is worth noting that since the EM adopts a globally uniform protection region diameter, then some regions can not ensure the lower bound of inference error, that is, not all regions satisfy \eqref{eq:inequality}, so that it can not preserve $\epsilon$-DP. To solve this, we use the maximum protection region diameter in DPIVE as the globally uniform diameter of protection region for EM ($13$km in GeoLife and $36$km in Gowalla, respectively). Then the quality losses of EM are $4.98$ and $16.26$ on the GeoLife and Gowalla, respectively, which are $1.7$ times  as large as those of DPIVE on average.

\begin{figure}[tb]
\centering
\subfigure[$ExpErr$ varing with $\epsilon$ (Geo.)]{
		\includegraphics[height=1.8 in]{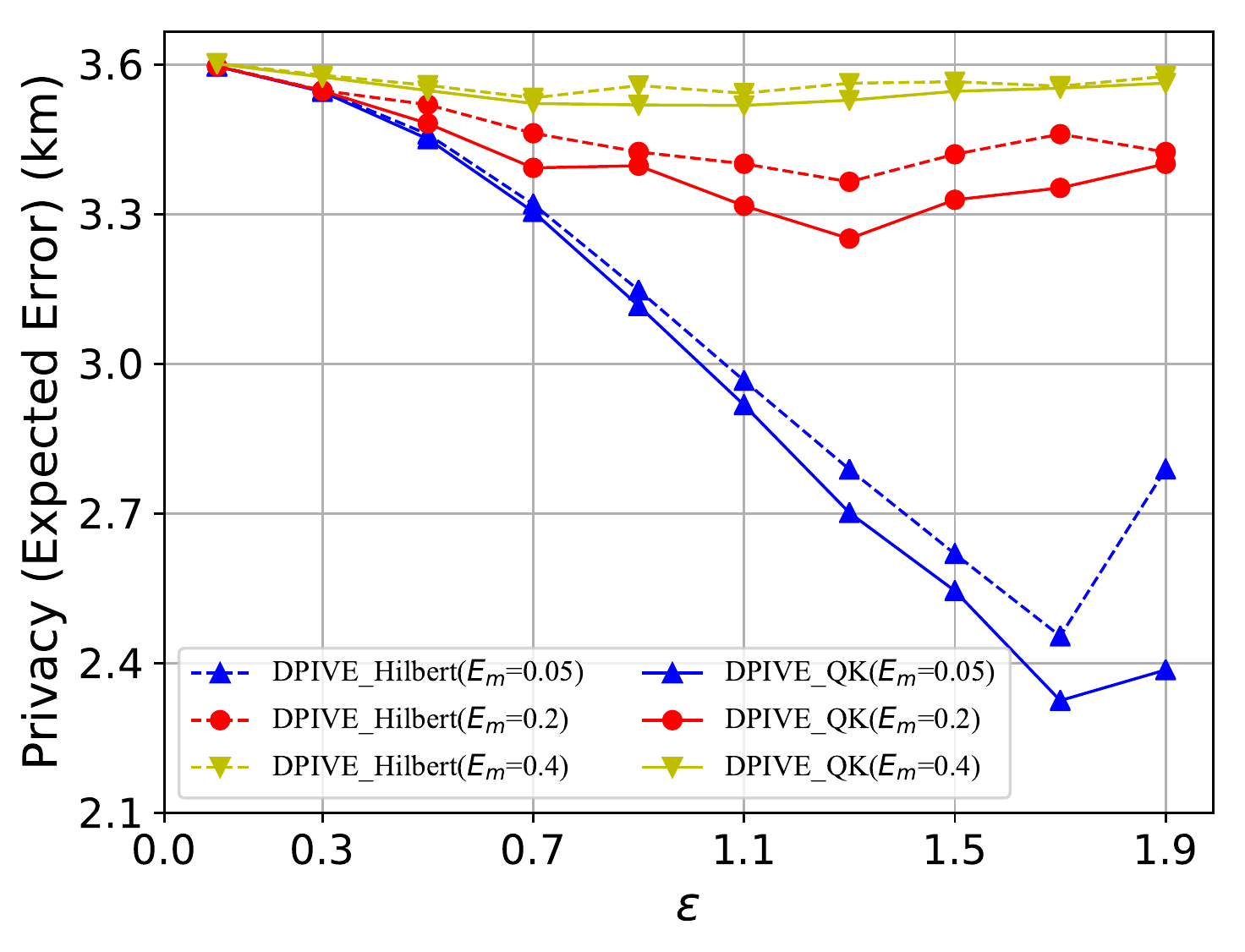}
	}
\subfigure[$QLoss$ varing with $\epsilon$ (Geo.)]{
		\includegraphics[height=1.8 in]{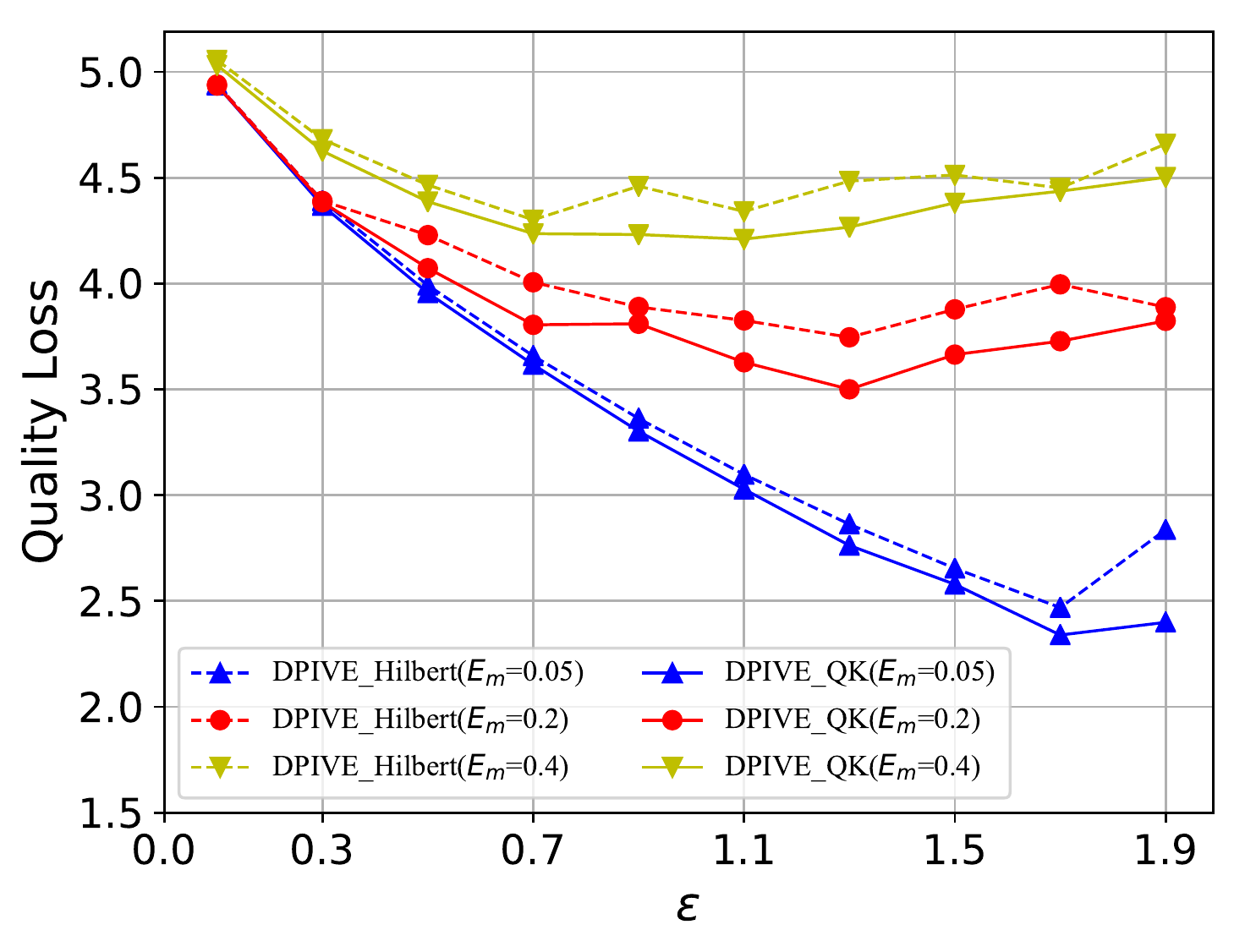}
	}
\subfigure[$ExpErr$ varing with $\epsilon$ (Gow.)]{
		\includegraphics[height=1.8 in]{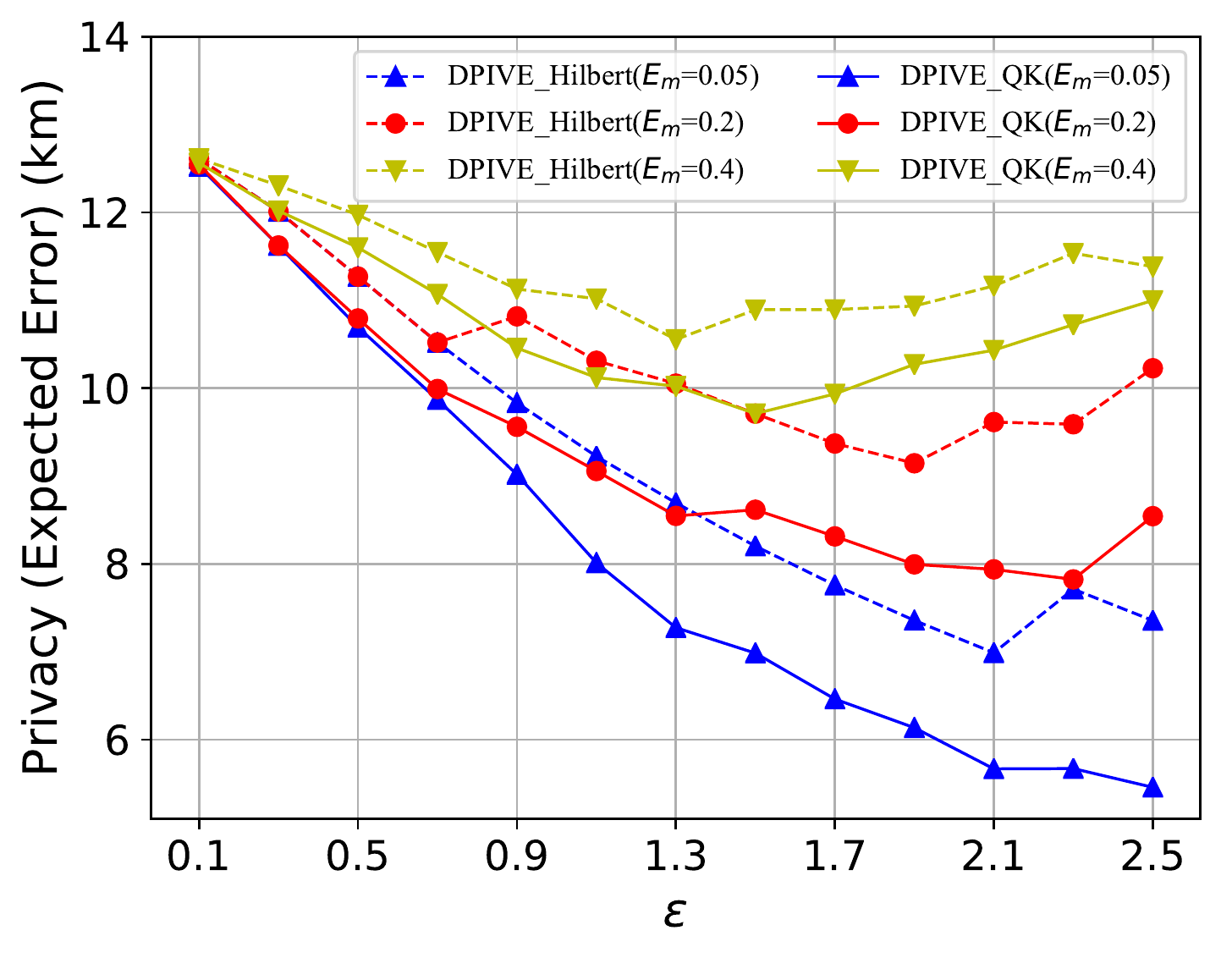}
	\label{fig:DPIVE_eps-Err}}
\subfigure[$QLoss$ varing with $\epsilon$ (Gow.)]{
		\includegraphics[height=1.8 in]{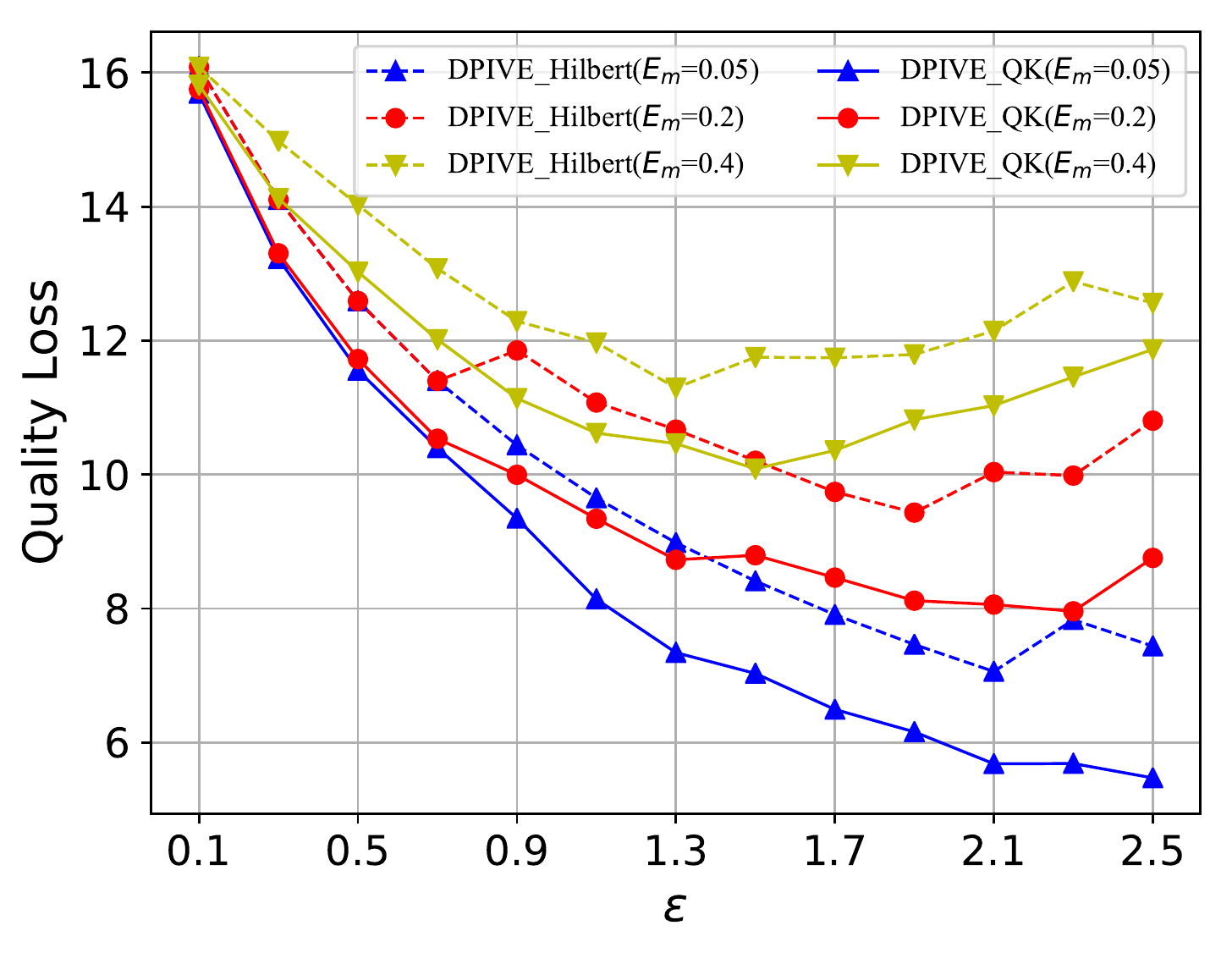}
	\label{fig:DPIVE_eps-QL}}
\caption{DPIVE\_Hilbert vs. DPIVE\_QK with varing $\epsilon$}
\label{fig:DPIVE_Joint_Eps}
\end{figure}

\begin{figure}[tb]
\centering
\subfigure[$ExpErr$ varing with $E_m$ (Geo.)]{
		\includegraphics[height=1.8 in]{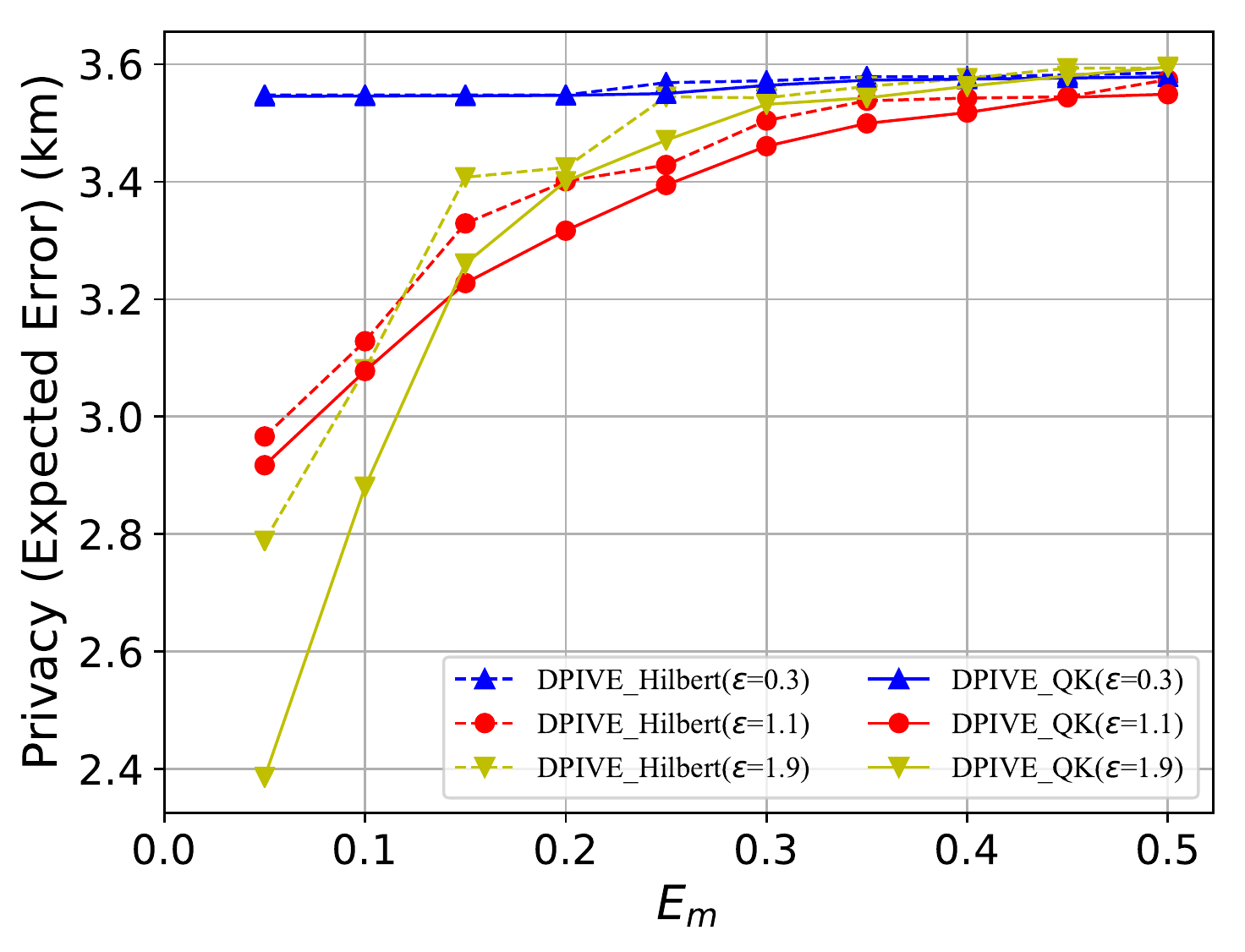}
	}
\subfigure[$QLoss$ varing with $E_m$ (Geo.)]{
		\includegraphics[height=1.8 in]{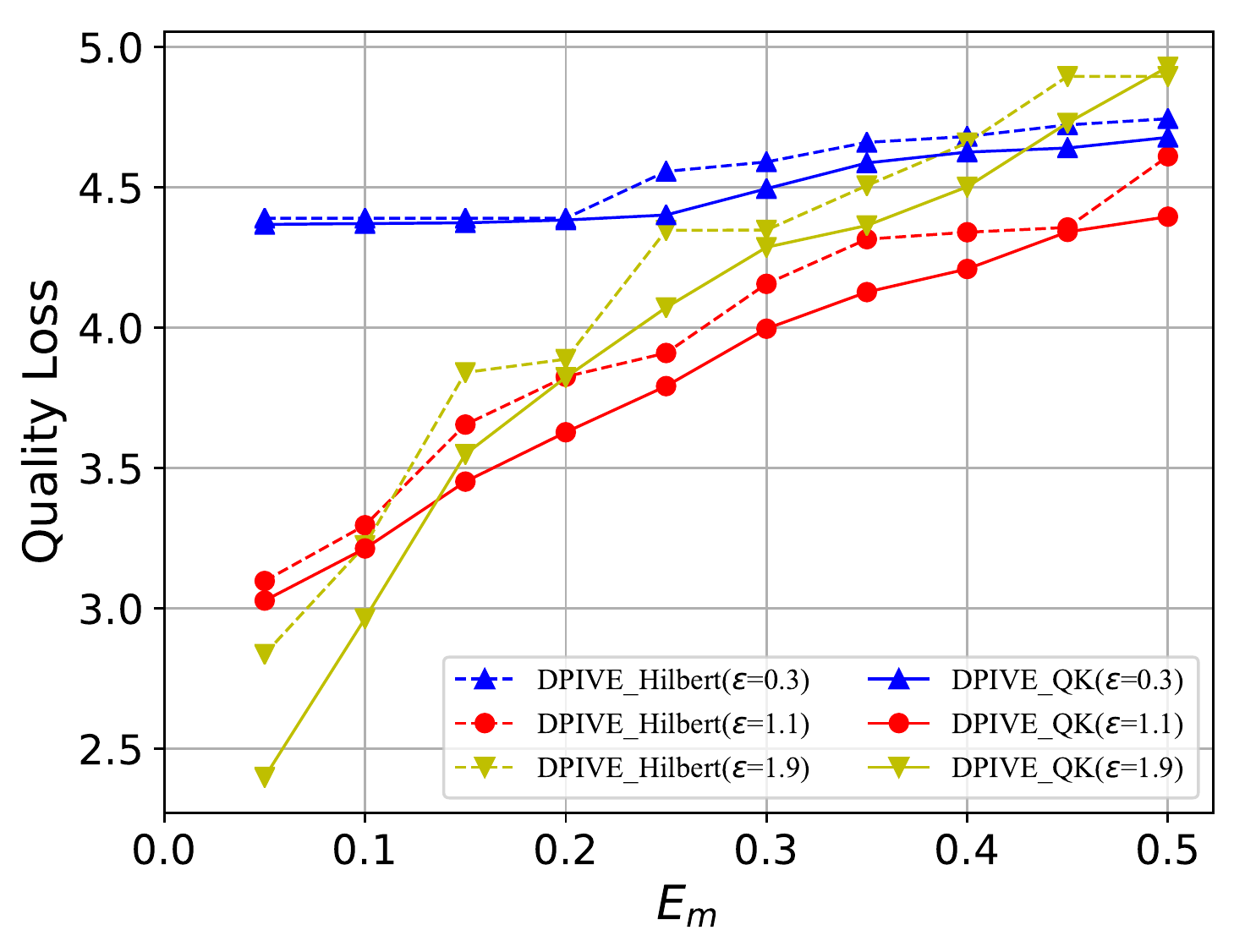}
	}
\subfigure[$ExpErr$ varing with $E_m$ (Gow.)]{
		\includegraphics[height=1.8 in]{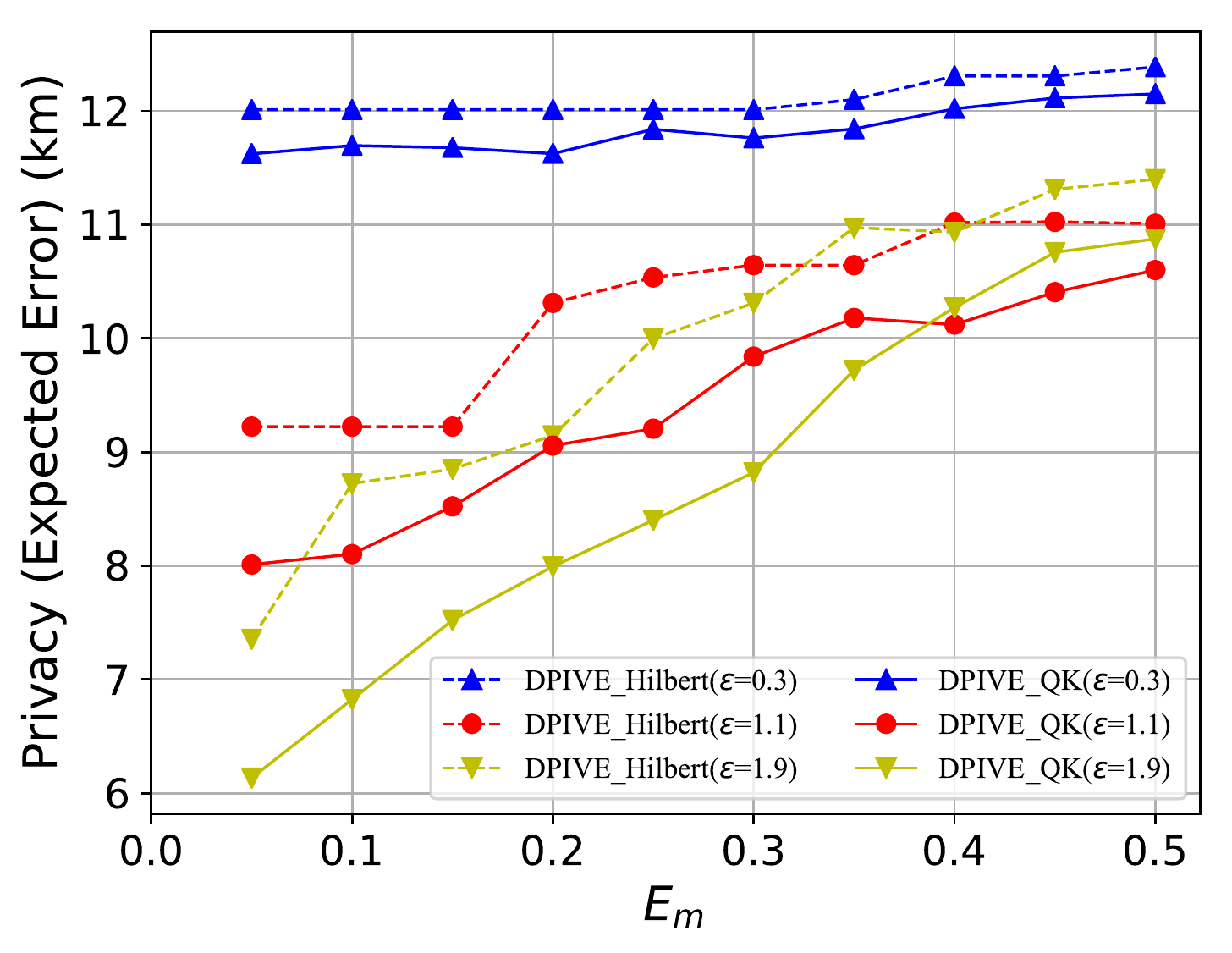}
	}
\subfigure[$QLoss$ varing with $E_m$ (Gow.)]{
		\includegraphics[height=1.8 in]{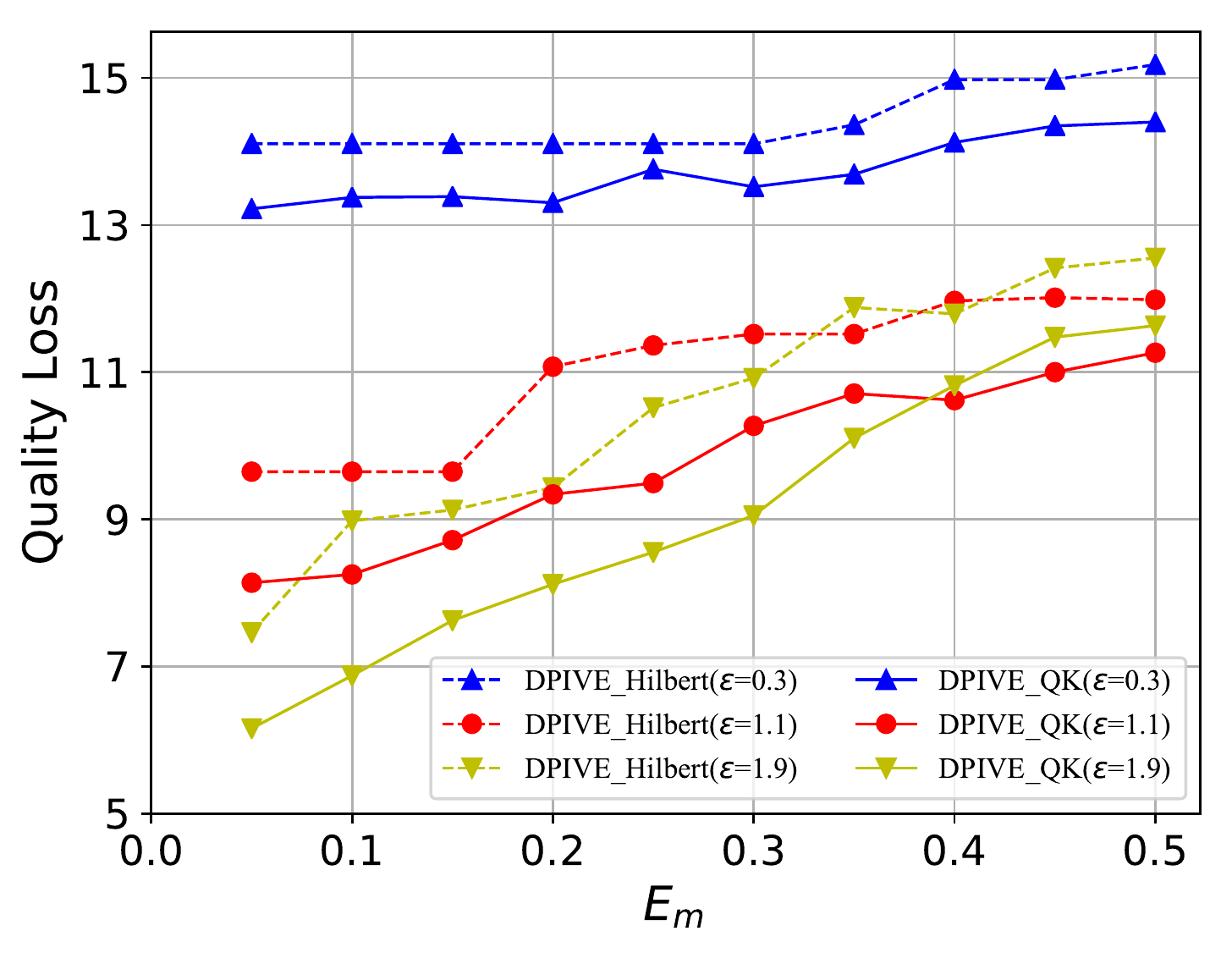}
	}
\caption{DPIVE\_Hilbert vs. DPIVE\_QK with varing $E_m$}
\label{fig:DPIVE_Joint_Em}
\end{figure}
\begin{figure}[tb]
\centering
\subfigure[$QLoss$ varing with $\epsilon$ (Geo.)]{
		\includegraphics[height=1.8 in]{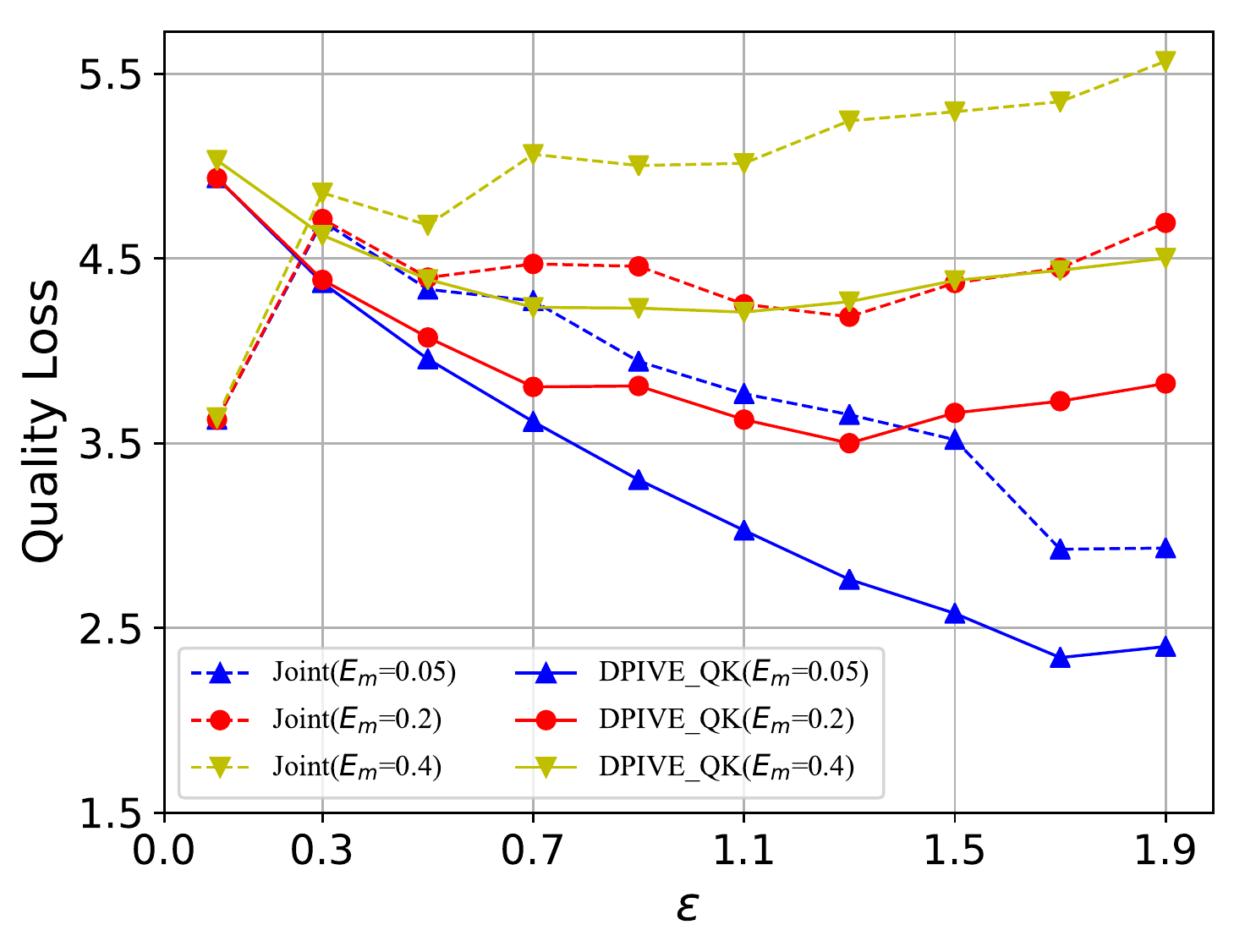}
	}
\subfigure[$QLoss$ varing with $E_m$ (Geo.)]{
		\includegraphics[height=1.8 in]{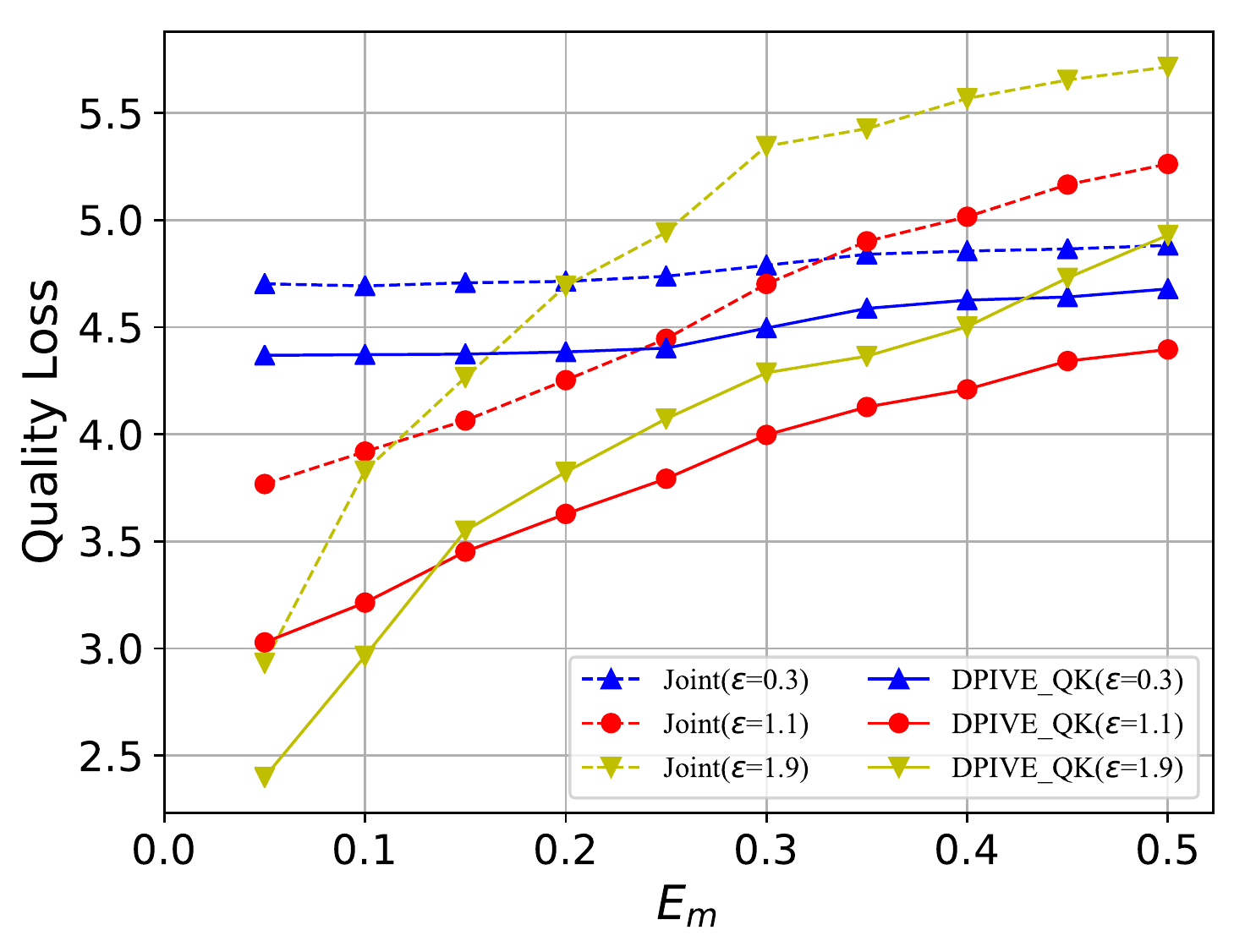}
	}
\subfigure[$QLoss$ varing with $\epsilon$ (Gow.)]{
		\includegraphics[height=1.8 in]{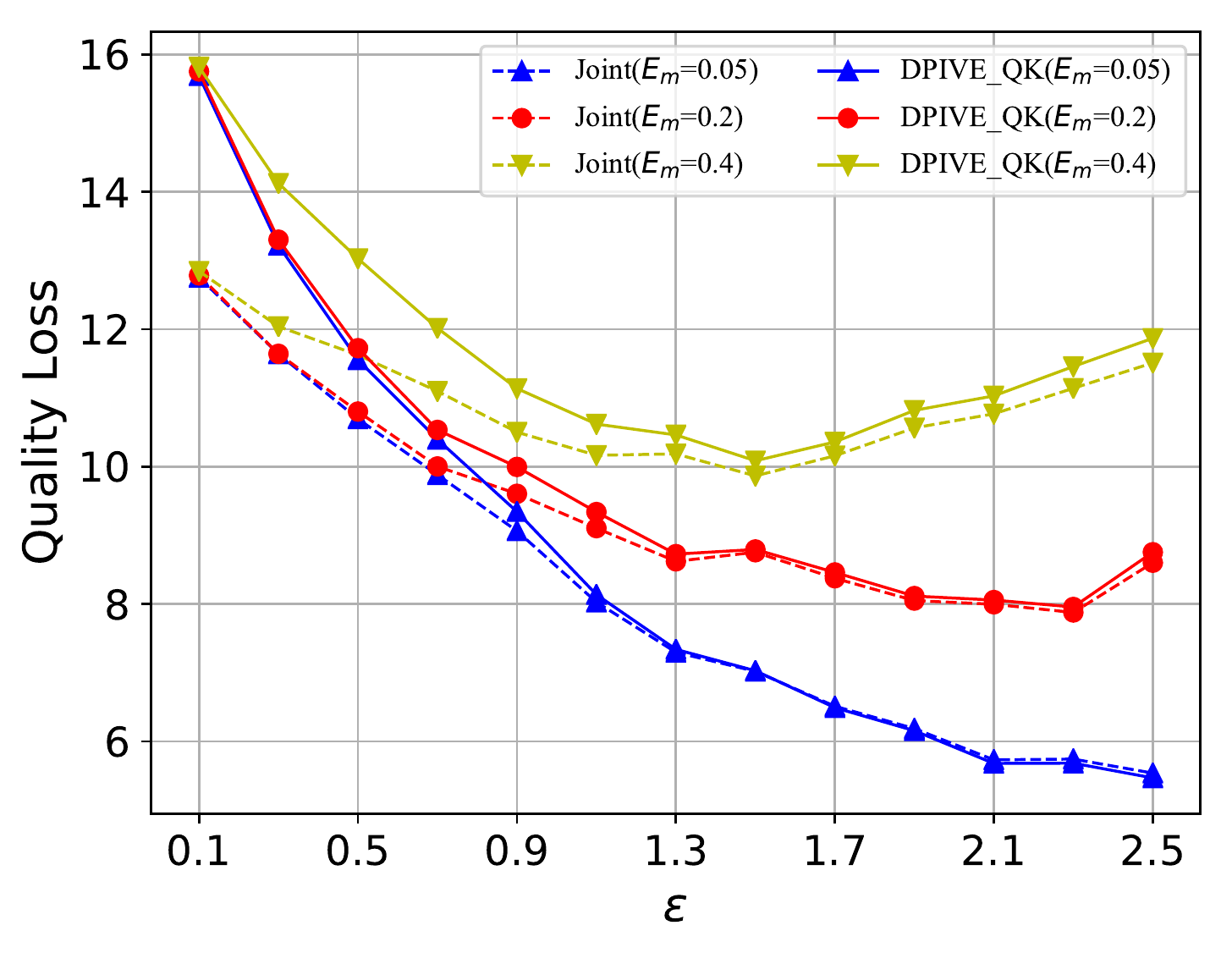}
	}
\subfigure[$QLoss$ varing with $E_m$ (Gow.)]{
		\includegraphics[height=1.8 in]{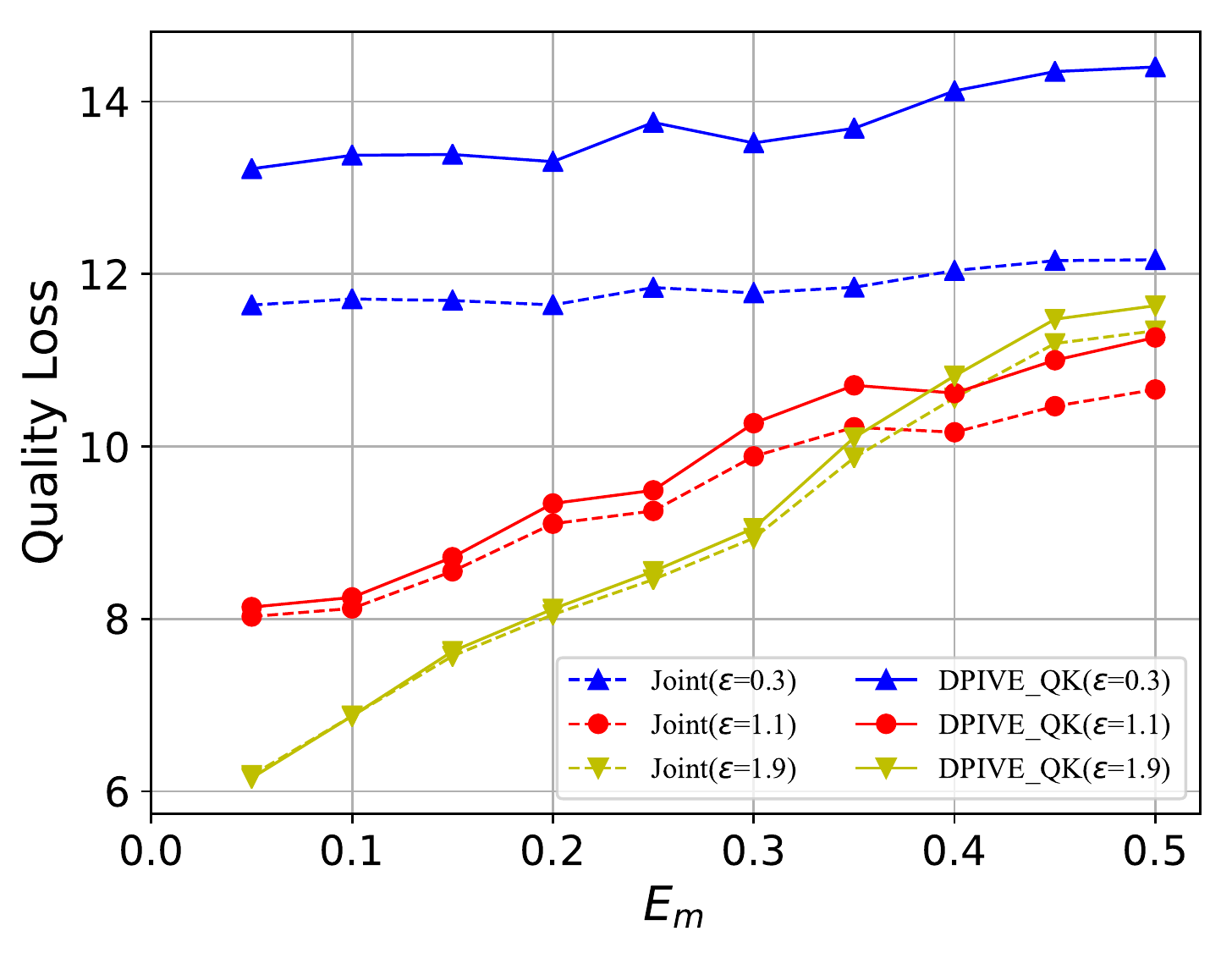}
	}
\caption{DPIVE\_QK vs. Joint on quality loss}
\label{fig:DPIVE_Joint}
\end{figure}

\textbf{Effect of QK-means.} Figs. \ref{fig:DPIVE_Joint_Eps} and \ref{fig:DPIVE_Joint_Em} show the comparison of the two approaches of DPIVE on the two datasets by changing $\epsilon$ and $E_m$. The results present similar trends on both datasets. The QK-means method reduces quality loss due to smaller diameter of PLSs as shown in Section \ref{sec:qkmeans}. When $e^{\epsilon}E_m$ is small, the two approaches almost coincide because almost all PLSs contain only two locations and the clustering method does not have obvious influence. In addition, Figs. \ref{fig:DPIVE_eps-Err} and \ref{fig:DPIVE_eps-QL} show that, as the privacy budget gradually increases, the quality loss for both approaches first decreases and then gradually increases. Indeed, when $\epsilon$ is large the diameters of PLSs increase rapidly, which brings greater quality losses.

\textbf{Comparing with Joint mechanism.} We compare the quality losses of DPIVE and Joint due to their combination of geo-indistinguishability and expected inference error. For convenience, DPIVE and Joint mechanisms are adjusted to have the same unconditional expected inference error. 
From Fig. \ref{fig:DPIVE_Joint}, the results show that DPIVE has lower utility loss in most cases in GeoLife, the quality loss of DPIVE is 9.7\% lower than that of Joint on average for $E_m=0.2$ while saving 15.8\% on average for $\epsilon=1.0$. In Gowalla, the quality losses are close for both schemes, while DPIVE provides better protection on skewed locations than Joint.

\subsection{Performance Analysis of PDPIVE}\label{subsect:PDPIVE}
\begin{figure}[tb]
\centering
\subfigure[Privacy vs. $E_m$ (Geo.)]{
		\includegraphics[height=1.8 in]{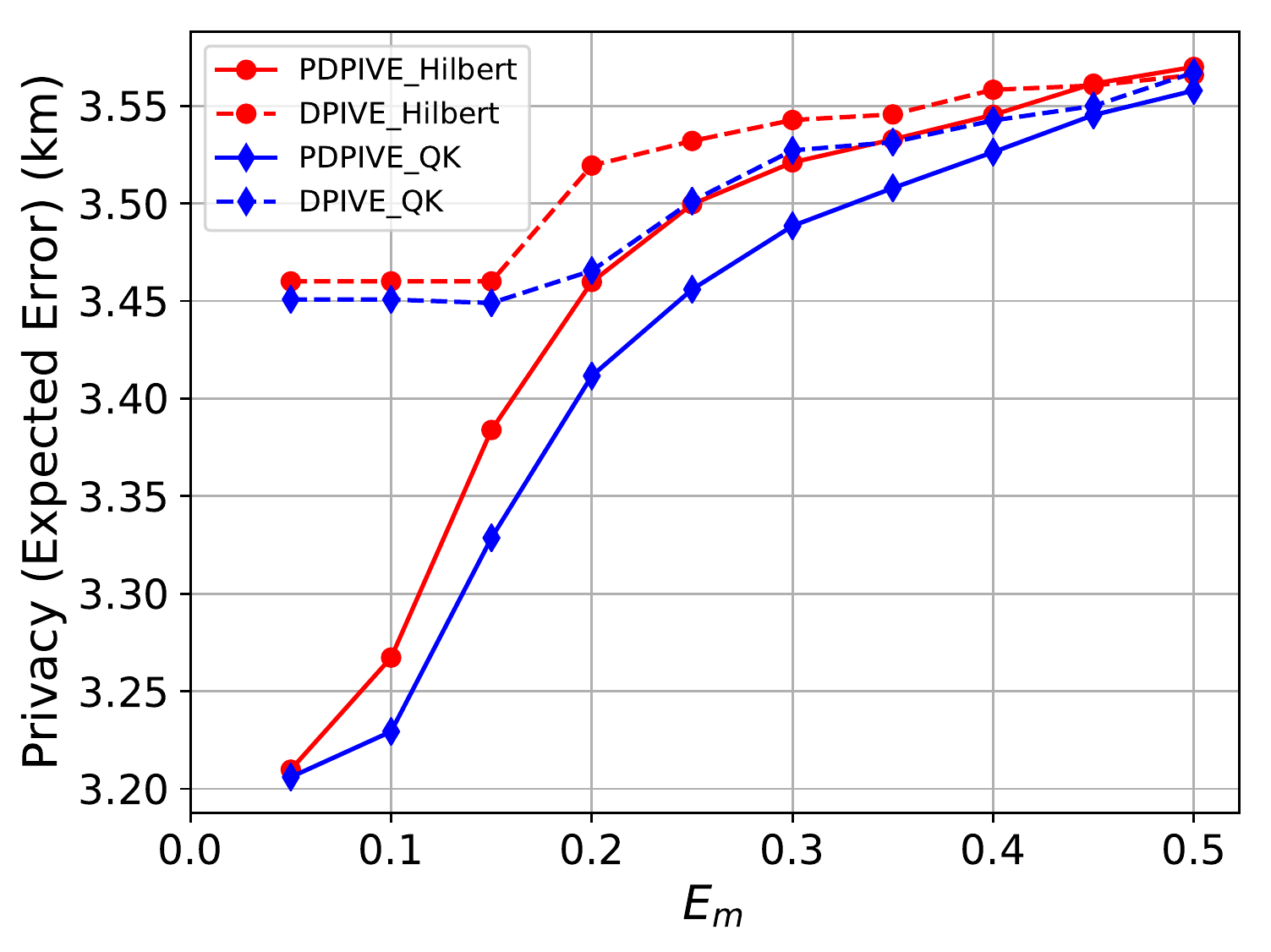}
	}
\subfigure[$QLoss$ vs. $E_m$ (Geo.)]{\label{subfig:em_loss_geo}
		\includegraphics[height=1.8 in]{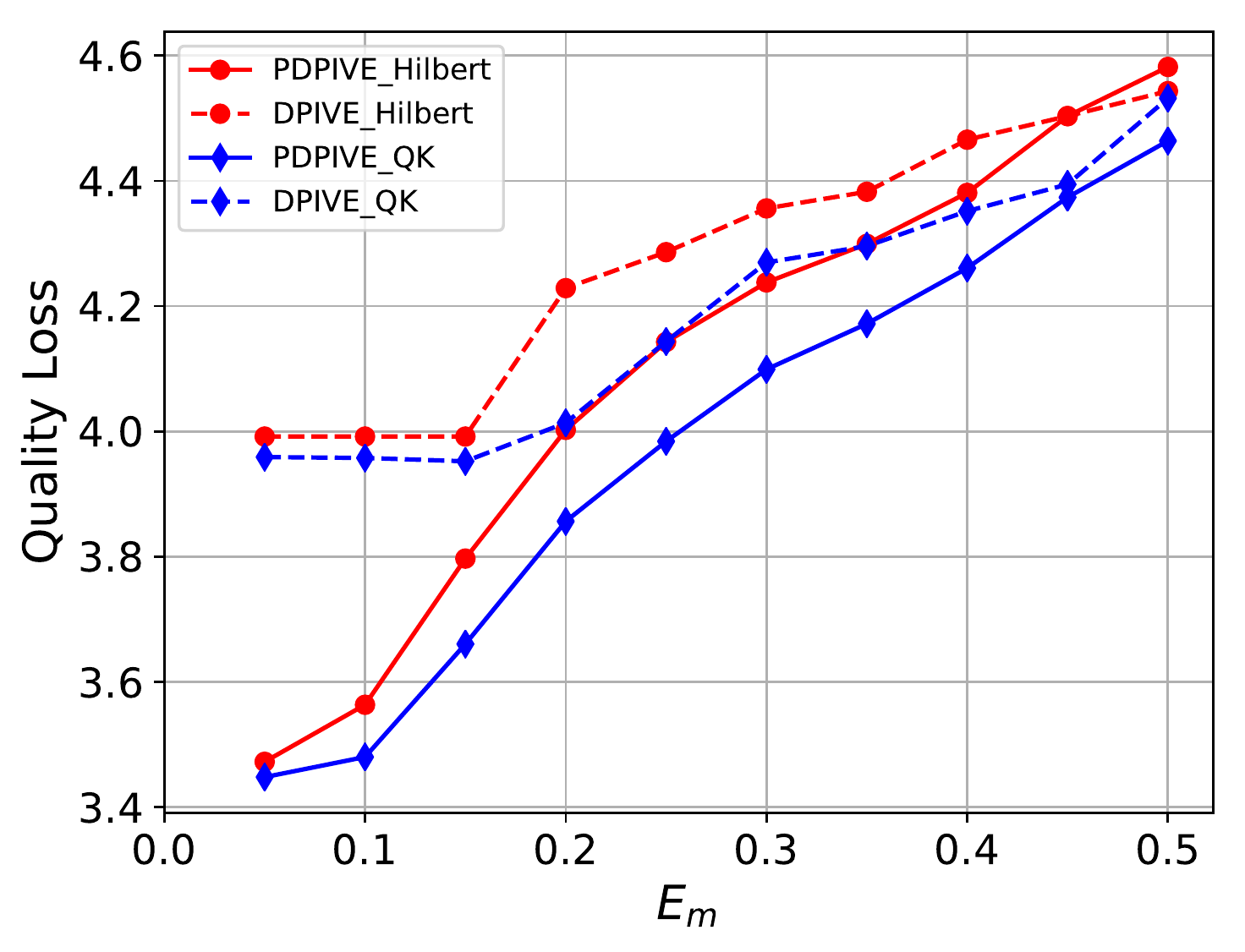}
	}
\subfigure[Privacy vs. $E_m$ (Gow.)]{
		\includegraphics[height=1.8 in]{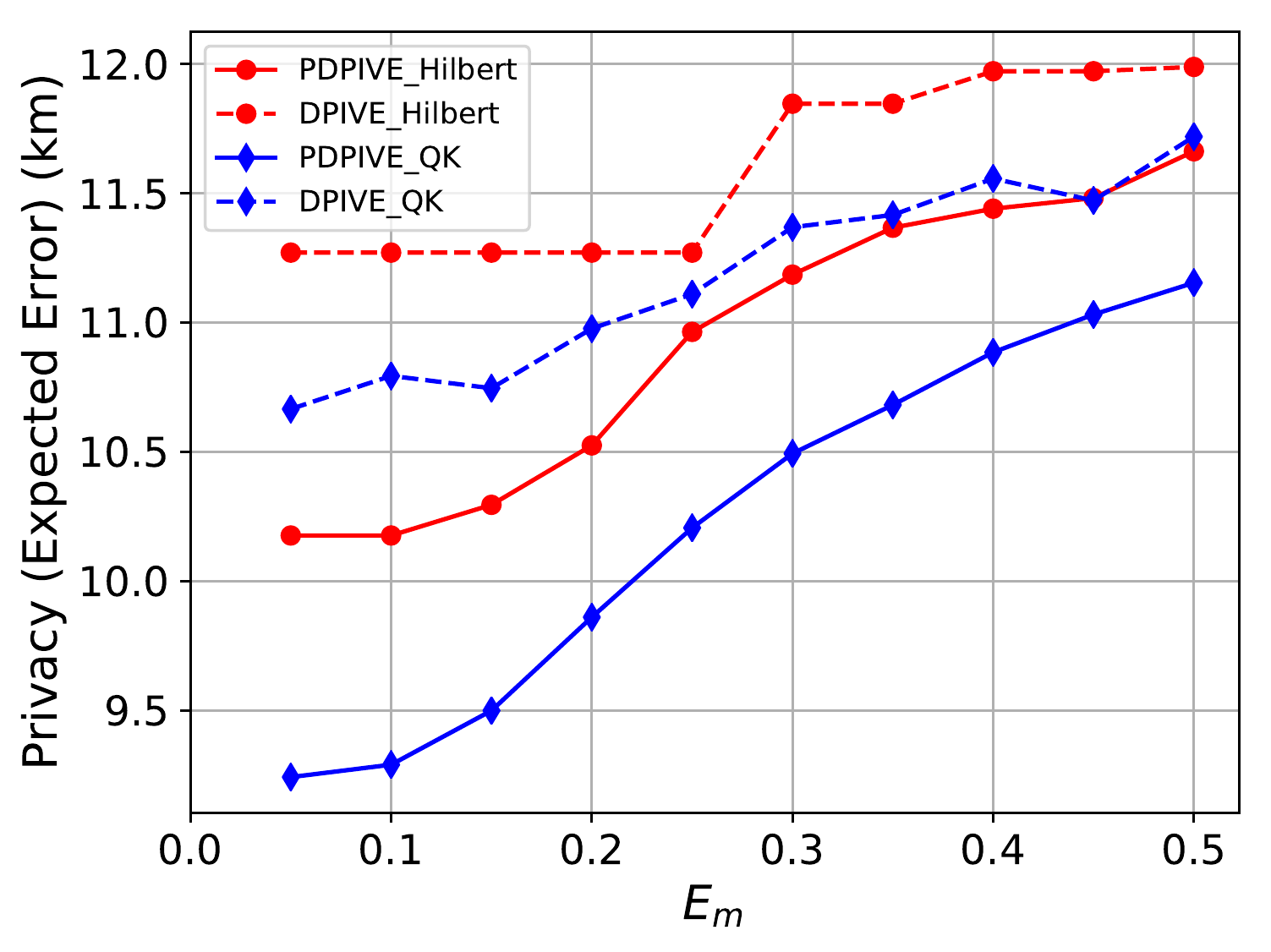}
	}
\subfigure[$QLoss$ vs. $E_m$ (Gow.)]{\label{subfig:em_loss_gow}
		\includegraphics[height=1.8 in]{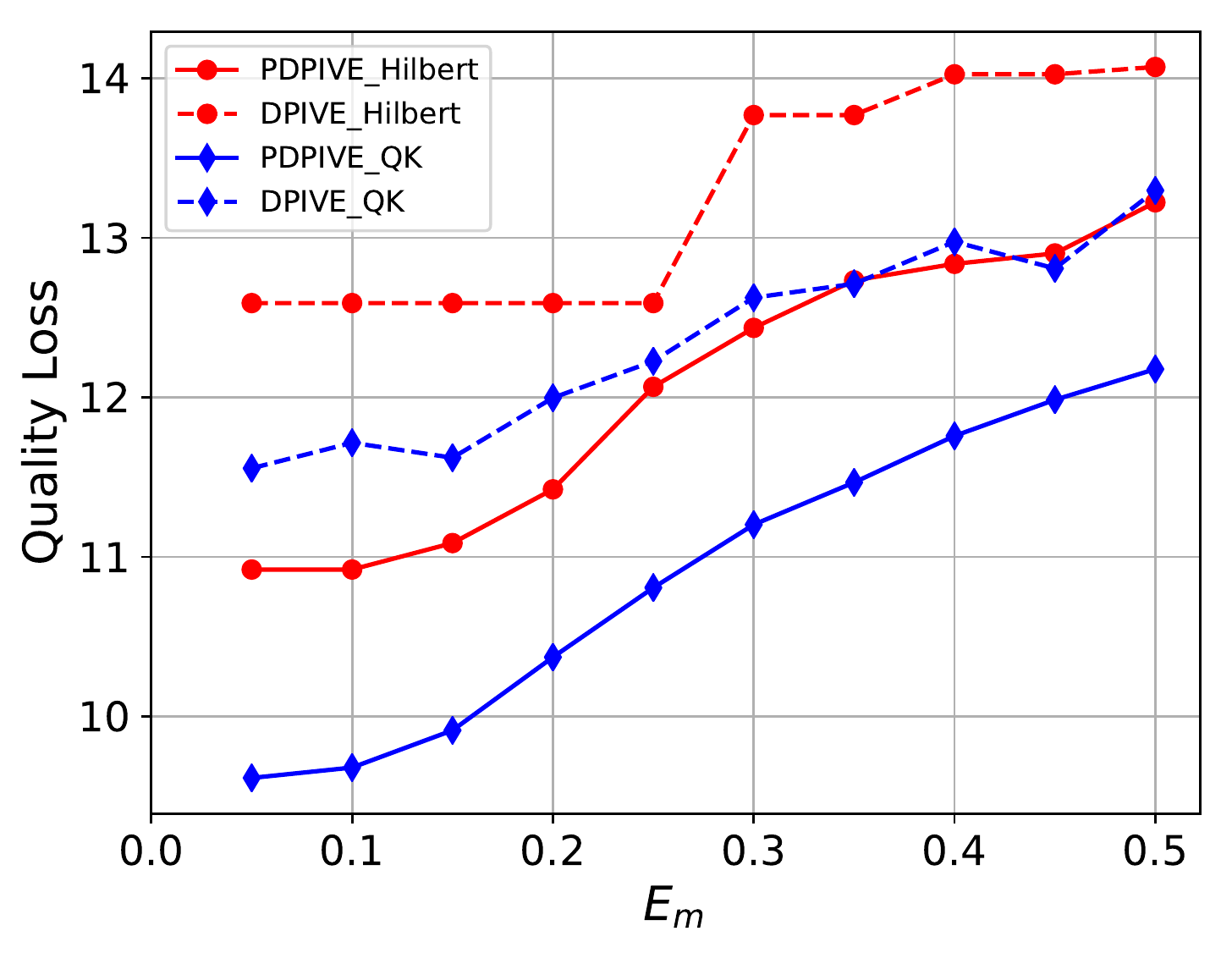}
	}
\caption{Performance of each scheme with personalized $\epsilon$.}
\label{fig:PDPIVE_ep}
\end{figure}

In this section, we mainly evaluate the impact of $\epsilon$'s personalization on the performance of PDPIVE. We focus on $QLoss$ for comparisons among four appoaches, two PDPIVE schemes (\emph{PDPIVE\_QK} and \emph{PDPIVE\_Hilbert}) and two DPIVE baselines (\emph{DPIVE\_QK} and \emph{DPIVE\_Hilbert}). To be specific, the personalized schemes, \emph{PDPIVE\_QK} and \emph{PDPIVE\_Hilbert}, search for optimal disjoint PLSs along respective lines as before, and each PLS meets the highest privacy requirements among the locations included while the baselines use the highest requirements in the whole $\mathcal{X}$. Besides, \emph{PDPIVE\_QK} constructs disjoint PLSs with considering the impact of weights \eqref{eq:w_ep}.
 The results of personalizing $\epsilon$ is shown in Fig. \ref{fig:PDPIVE_ep}. Our analysis is given from two perspectives.

Compared with the baselines, personalized schemes can effectively reduce quality loss. With personalized $\epsilon$,  the schemes, \emph{PDPIVE\_Hilbert} and \emph{PDPIVE\_QK}, reduce quality loss by $4.1\%$ and $4.9\%$ on GeoLife, and $9.1\%$ and $11.8\%$ on Gowalla, respectively (Fig. \ref{fig:PDPIVE_ep}). Since the baseline schemes adopts globally unified privacy parameters that meet the highest privacy requirements, many regions are protected with privacy level much higher than their requirements, which results in greater quality losses.

In terms of region partitioning strategy, compared with Hilbert method, QK-means method has lower quality loss. With personalizing $\epsilon$, the quality loss are reduced by an average of $2.9\%$ and $9.6\%$ on two datasets, respectively (Fig. \ref{fig:PDPIVE_ep}). Obviously, QK-means method has more advantages on Gowalla, which
 is mainly due to the fact that Gowalla locations are sparser than those in GeoLife and has more selection space in clustering. Although the privacy level of QK-means method  declines to some extent, it satisfies privacy requirements in each region.

 \subsection{Application Analysis}

In this section, we make an application analysis in terms of Spatial Crowdsourcing (SC) \cite{WYH19, WWY22}. The workers send their false locations to the SC-Server, which assigns each task to the nearest three idle workers according to the reported locations after receiving the task request. The metric WTD stands for the average distance that the reported workers travels from the actual location to the allocated task. This reflects the efficiency of mechanism application and measures the service availability to a certain extent \cite{LHX06,XHL13}.

We conduct comparative experiments on the two datasets with varying privacy parameter $\epsilon$. Under each parameter setting, we sample randomly $100$ single-tasks in each dataset (with 30 idle workers, respectively) and average their WTDs. Any two tasks are assumed to have no spatio-temporal confliction to each other so that they can share a single worker.
The notation Non-privacy means DPIVE without privacy protection, that is, the SC-server geocasts the three idle workers closest to the task directly based on the real locations and their average WTD is referred to.

\begin{figure}[tb]
\centering
\includegraphics[width=0.40\linewidth]{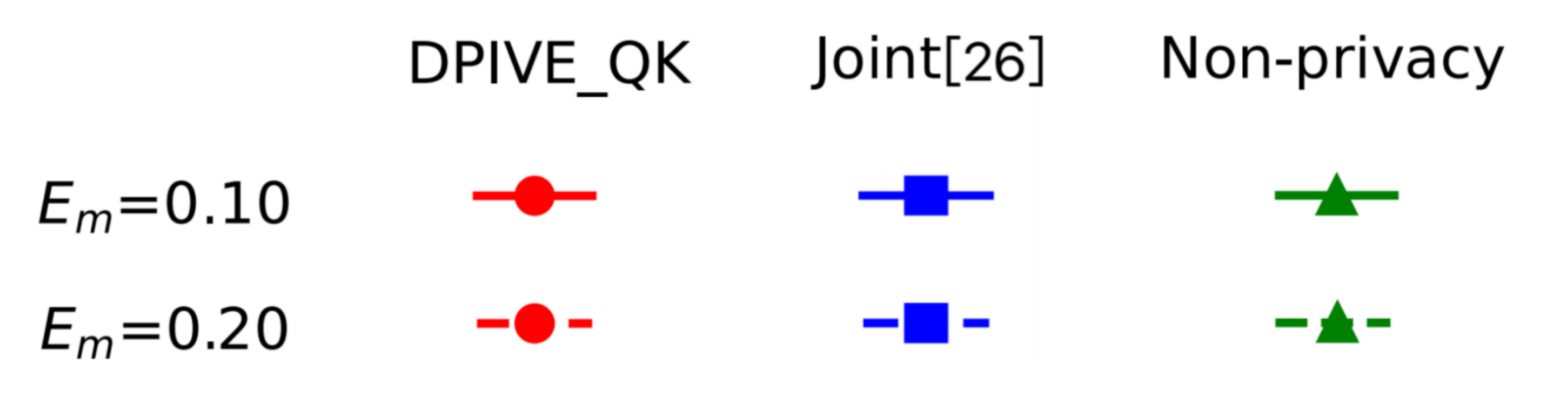}\\
\subfigure[WTD, GeoLife]{
		\includegraphics[height=1.8 in]{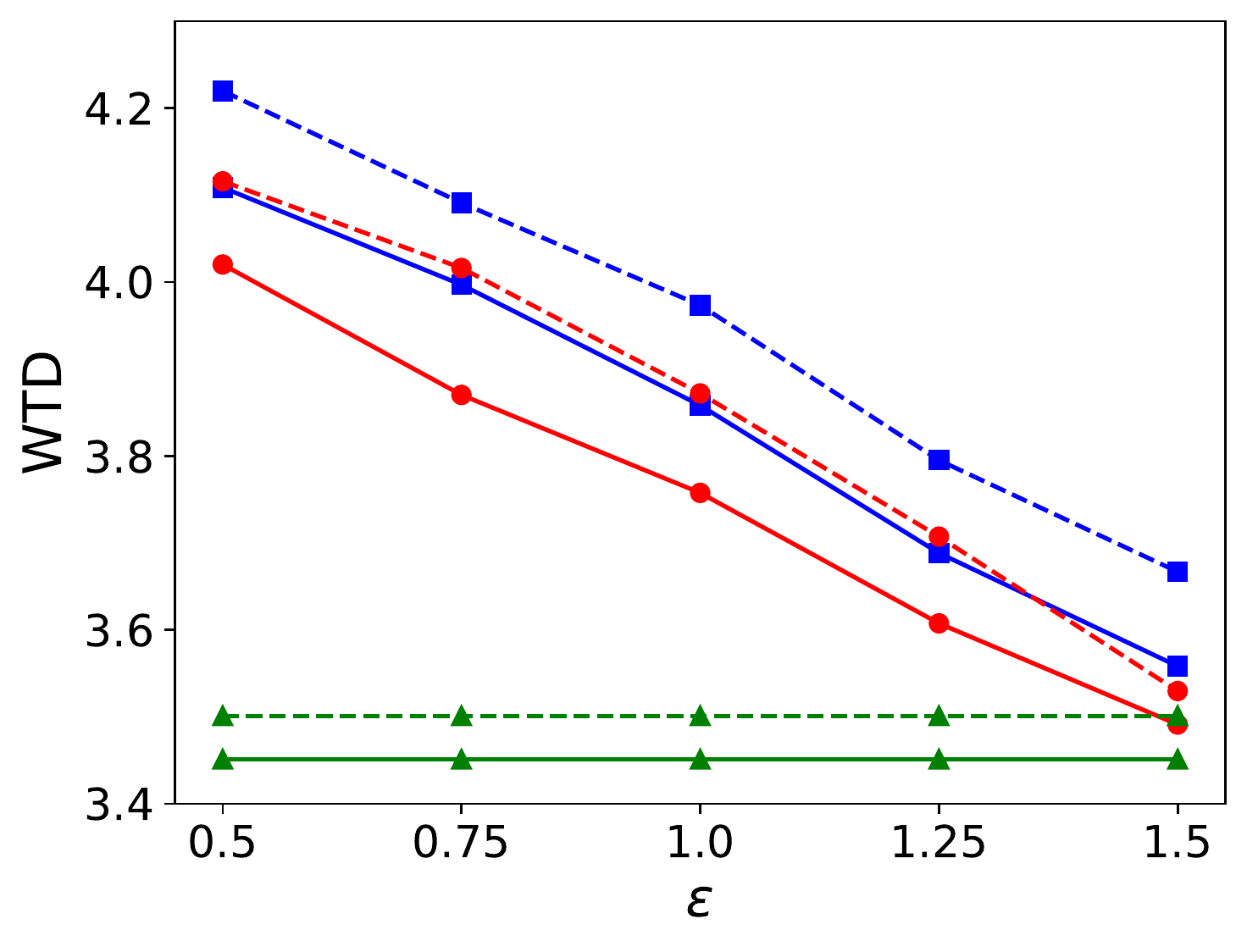}
	}
\subfigure[WTD, Gowalla]{
		\includegraphics[height=1.8 in]{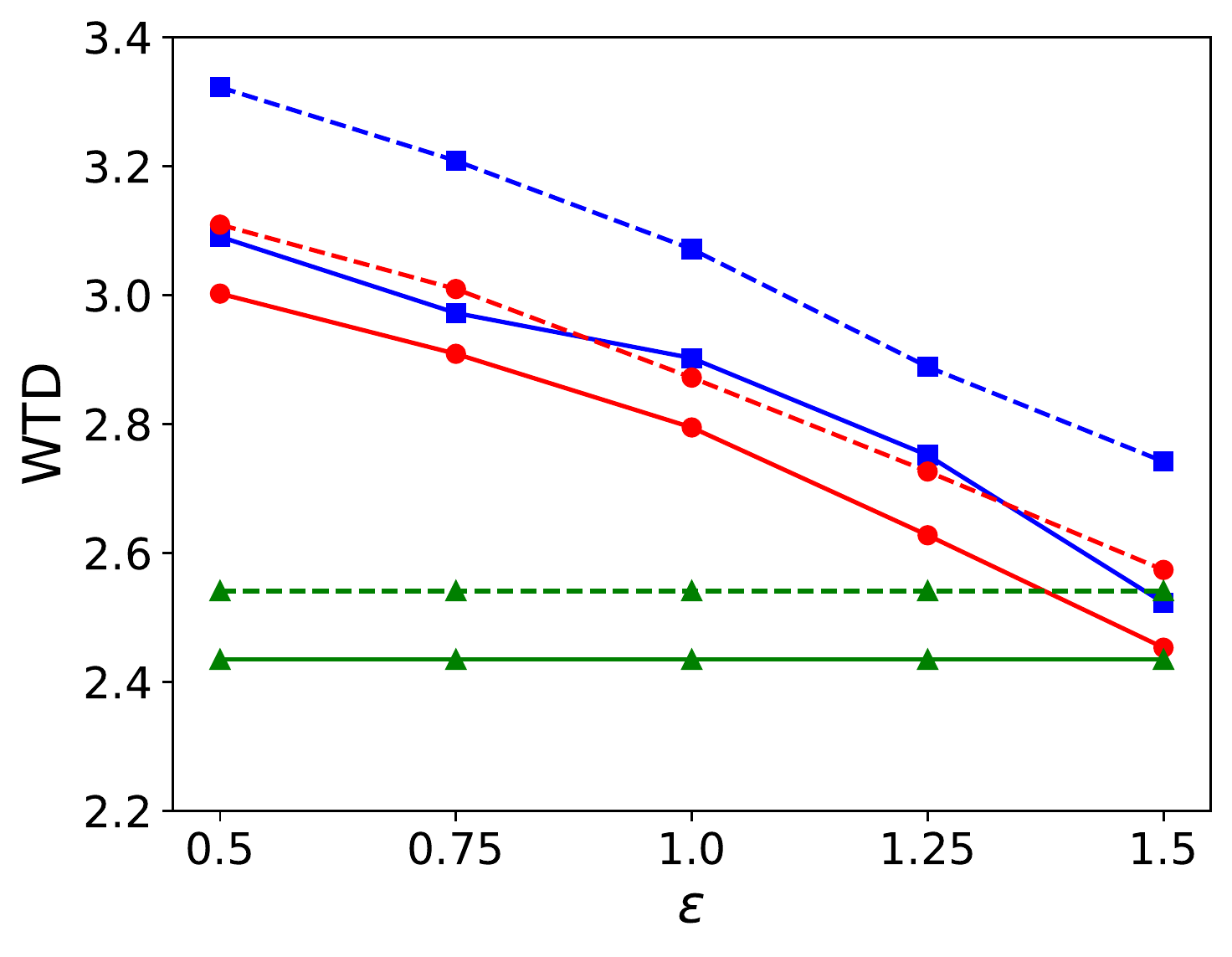}
	}
\caption{Comparisons on WTD with varying $\epsilon$.}
\label{fig:PDPIVE_WTD}
\end{figure}

Fig. \ref{fig:PDPIVE_WTD} shows that compared to DPIVE, Joint \cite{Sho15} has an average increase of $2.5\%$ and $3.3\%$, and a maximum increase of $3.2\%$ and $4.7\%$, respectively, on the two datasets at $E_m=0.10$, while giving an average increase of $2.7\%$ and $6.6\%$ and a maximum increase of $3.9\%$ and $7.0\%$, respectively, at $E_m=0.20$.
This shows that our mechanism can improve the availability of existing SC mechanisms while guaranteeing the protection level of worker location privacy.

\section{Conclusions and future work}\label{sec:conclusion}

This paper investigates the differential privacy preservation of location obfuscation mechanism based on problematic PIVE framework. Since PIVE fails to offer differential privacy guarantees on adaptive Protection Location Set (PLS), we develop DPIVE, a regionalized location obfuscation mechanism. According to the relevant privacy parameters and their relationship, the entire location set is partitioned into multiple disjoint PLSs, and the locations in the same PLS share the same sensitivity of utility. Each PLS satisfies the lower bound of the inference error for the locations inside. The apriori locations within the same PLS are strongly geo-indistinguishable to each other, while those locations across different PLSs satisfy weak differential privacy. As a generalization that allows users to personalize their own privacy levels, we first design a quasi $k$-means clustering algorithm and implement the location obfuscation mechanism PDPIVE theoretically and practically. Experiments with two public datasets demonstrate that our mechanisms improve
significantly the performance, particularly on skewed locations. In the future work, we will explore differential location privacy problems in the large-scale domain scenario with applications, which involves higher computational complexity and various requirements on communication environments, such as the forthcoming paper \cite{ZZLC2022}.

\vspace{5mm}

\bibliographystyle{ACM-Reference-Format}
\bibliography{DPIVE}

\end{document}